\documentclass[apj]{emulateapj}
\usepackage{amsmath}
\usepackage{natbib}
\newcommand{\HeI}{\hbox{{\rm He}\kern 0.1em{\sc i}}}
\newcommand{\HeII}{\hbox{{\rm He}\kern 0.1em{\sc ii}}}
\newcommand{\CIV}{\hbox{{\rm C}\kern 0.1em{\sc iv}}}
\newcommand{\NIII}{\hbox{{\rm N}\kern 0.1em{\sc iii}}}

\shorttitle{Young Stellar Clusters}
\shortauthors{Leistra et al.} 

\begin{document}

\title{Two 2MASS-Selected Young Stellar Clusters:  Photometry, Spectroscopy,
  and the IMF}
\author{A.~Leistra\altaffilmark{1}, A.~S.~Cotera\altaffilmark{2},
  J.~Liebert\altaffilmark{1}, \& M.~Burton\altaffilmark{3}}
\altaffiltext{1}{Steward Observatory, University of Arizona, 933 N. Cherry 
Ave., Tucson, AZ 85721}
\altaffiltext{2}{SETI Institute, 515 N. Whisman Road, Mountain View, CA 94043}
\altaffiltext{3}{School of Physics, University of New South Wales, Sydney 2052,
  NSW, Australia}
\email{aleistra@as.arizona.edu, acotera@seti.org, jliebert@as.arizona.edu, mgb@phys.unsw.edu.au}

\begin{abstract}

We present near-infrared $J$, $H$, and $K_s$ images and $K$-band spectroscopy of
two newly discovered stellar clusters at different stages of evolution. Our
spectra suggest the presence of massive YSOs in the heavily embedded cluster in
the star-forming region near radio source G353.4-0.4 and an O5-O6V star in the
cluster near radio source G305+00.2.  We determine a $K$-band luminosity
function (KLF) for both clusters and an initial mass function (IMF) for the
cluster near G305+00.2.  The derived IMF slope is $\Gamma = -1.5$ if the KLF
is used to derive the IMF and is $\Gamma = -0.98$ if the color-magnitude
diagram and spectra are used.  The more reliable CMD-based slope is flatter
than the Salpeter value usually found for stellar clusters.  We find that using 
the KLF alone to derive an IMF is likely to produce an overly steep slope in 
stellar clusters subject to variable extinction.
 
\end{abstract}


\section{INTRODUCTION}
\label{sec:intro}

Despite their intrinsic rarity and short lifetimes, massive stars are extremely
important in the evolution of galaxies.  They play an important role in
determining the course of the formation of less massive stars, though the 
nature of this role is still uncertain, and their stellar winds and eventual
supernovae shape the interstellar medium.  They produce most of the heavy
elements in the universe, as well as much of the UV radiation in galaxies.
Their rarity, combined with the effects of large Galactic extinctions, often
results in the availability of more comprehensive studies of massive stars in
external galaxies, where the entire stellar population can be observed at once,
than within our own where massive stars must be studied individually and the
census of massive stars is still very incomplete.  High optical extinction
within the galactic plane ($A_V\gtrsim 20$) has limited optical studies of
massive stars to relatively nearby regions \citep[R$_{\rm{solar}}\lesssim3.0$
kpc,][]{massey03}.  Even within that radius, optically selected catalogs of O
stars have been found to be incomplete, especially in star-forming regions and
young clusters \citep[e.g.][]{hanson95}.  This incompleteness necessitates the
use of infrared, radio and X-ray observations, particularly in the inner
regions of the Galaxy and in star formation regions.  The near-infrared (NIR,
1-5~\micron) is an especially useful regime for the study of massive stars; the
stellar atmosphere is still observed directly, but since for example $A_K
\simeq 0.11 A_V$, we can observe these stars in regions where dust, either
along the line of sight or local to the star-forming region, makes them
inaccessible at optical wavelengths.  The discovery and characterization of
stellar clusters observable only in the infrared can significantly enhance our
understanding of obscured Galactic regions which harbor embedded massive stars
or massive protostars.
  
Recent studies indicate that clusters may account for 70-90\% of star formation
and that embedded clusters (those still partially or fully enshrouded in their
natal molecular cloud) may exceed the number of more traditional open clusters
by a factor of $\sim$20 \citep{elmegreen00,lada03}.  In the last decade,
advancements in NIR observational capabilities resulted in the discovery and
classification of some of the most massive young stellar clusters in the
Galaxy, each containing dozens of O and WR stars
\citep[e.g.][]{nagata95,coteraarches,figer96}.  Recent studies \citep{figer99}
have suggested that within these clusters, the initial mass function (IMF) does
not follow the canonical Salpeter form with a slope $\Gamma = -1.35$, but
instead is more heavily weighted toward massive stars; mass segregation has
been proposed as a solution \citep{stolte02}.  In the last several years a
number of studies of well-known star formation regions have also been carried
out in the NIR, \citep[e.g][]{okumura00,blum01,blum02,figueredo02}.  These
studies have in most cases found an IMF consistent with the Salpeter value, and
have uncovered candidate massive YSOs.  In addition, within the past ten years,
massive YSOs within molecular clouds have been studied in the NIR,
\citep[e.g.][]{chakrabotry,ishii} and in young stellar clusters
\citep[e.g.][]{hanson97}. Massive YSOs, however, remain significantly less
studied and are poorly understood in comparison with their lower-mass
counterparts; many more must be identified and studied before we can adequately
address how the formation of massive stars differs from that of low-mass stars.

The final release of the Two Micron All Sky Survey (2MASS) has fostered studies
which can probe the entire Galaxy for previously unknown stellar clusters.
Initial attempts were made which searched for stellar density enhancements,
\citep[e.g.][]{db00,db01,db03}, but the identification of previously unknown
clusters has met with limited success.  For example, \citet{db00} identified 52
candidate clusters, which subsequent observations \citep{db03} indicated were
in fact 10 confirmed clusters, 3 ``probable'' clusters, and 11 ``dissolving
cluster candidates''; the remainder were not clusters.  Our observations of at
least one of the \citet{db03} ``confirmed clusters'', however, indicates that
the ``cluster'' is most likely a region of low extinction rather than a true
cluster \citep{cotera05}. We have performed an independent search of the 2MASS
archive, using color criteria in addition to stellar density enhancements.  We
have searched in the vicinity of regions identified as likely sites of star
formation based on radio and IRAS far-infrared flux ratios, and are currently
conducting a search of the entire 2MASS Point Source Catalog.  We search the
Point Source Catalog for regions of higher stellar density than the background
(determined locally within a 5\arcmin radius) which are redder in $H-K$ than
the local field.  This selects for embedded clusters, with the color criteria
helping to eliminate chance superpositions and regions of low extinction.  In
contrast, Dutra \& Bica (2000,2001) use only stellar density to select
clusters.  Our method has been relatively successful to date; correctly
selecting 7 clusters out of 9 potential targets, including 4 candidates toward
the inner Galaxy. We present NIR imaging and spectroscopy of the two confirmed
clusters in the inner Galaxy in this paper, and discuss the two unconfirmed
targets in detail in \citet{cotera05}.  The cluster near G305.3+0.2 was
independently discovered by \citet{db03a}.  The additional 5 outer-galaxy
targets are described in Paper II.

NIR imaging and spectroscopy of both young stellar clusters and nascent stellar
clusters enables us to expand the study of the IMF in objects where there has
been little to no stellar evolution off the main sequence or cluster
evaporation, and where the cluster age can be constrained to within $\sim 2$ Myr.
Spectral typing of the most massive stars in the cluster allows their masses to
be determined relatively precisely, and when combined with photometry it
facilitates a reliable determination of the masses of stars throughout the
entire cluster (Massey, Johnson, \& DeGioia-Eastwood 1995; \citealt{massey02}),
allowing the initial mass function of the cluster to be determined more
accurately than photometry alone would permit.  In this paper we present the
results of NIR observations of two clusters found toward the inner Galaxy,
which we designate by the Galactic coordinates of their centers, G353.4-0.36
(17:30:28 -34:41:36 J2000) and G305.3+0.2 (13:11:39.6 -62:33:13 J2000).  In
Paper II we will present the results of similar observations of five clusters
in the outer Galaxy.

In \S\ref{sec:obs} we present the observations and data reduction, in
\S\ref{sec:results} we present the spectra and classifications of the
spectroscopically observed cluster members as well as the color-magnitude
diagrams, and in \S\ref{sec:klfimf} we describe the luminosity function and the
initial mass function.

\section{OBSERVATIONS \& DATA REDUCTION}
\label{sec:obs}

We observed candidate young stellar clusters with the facility instrument IRIS2
on the 3.9m Anglo-Australian Telescope (AAT) on July 12-15, 2003.  IRIS2 is an
imaging spectrometer which uses a 1024x1024 Rockwell HAWAII-1 HgCdTe array with
a platescale of 0\farcs45/pixel, resulting in a 7\farcm7$\times$7\farcm7 field
of view.  Images were obtained in $J$ (1.25~\micron), $H$ (1.63~\micron), and
$K_s$ (2.14~\micron) filters.  $R\simeq2300$ spectra of selected stars in each
cluster candidate were obtained in $K$ for all candidates.

\begin{figure*}
\scalebox{0.8}{\includegraphics{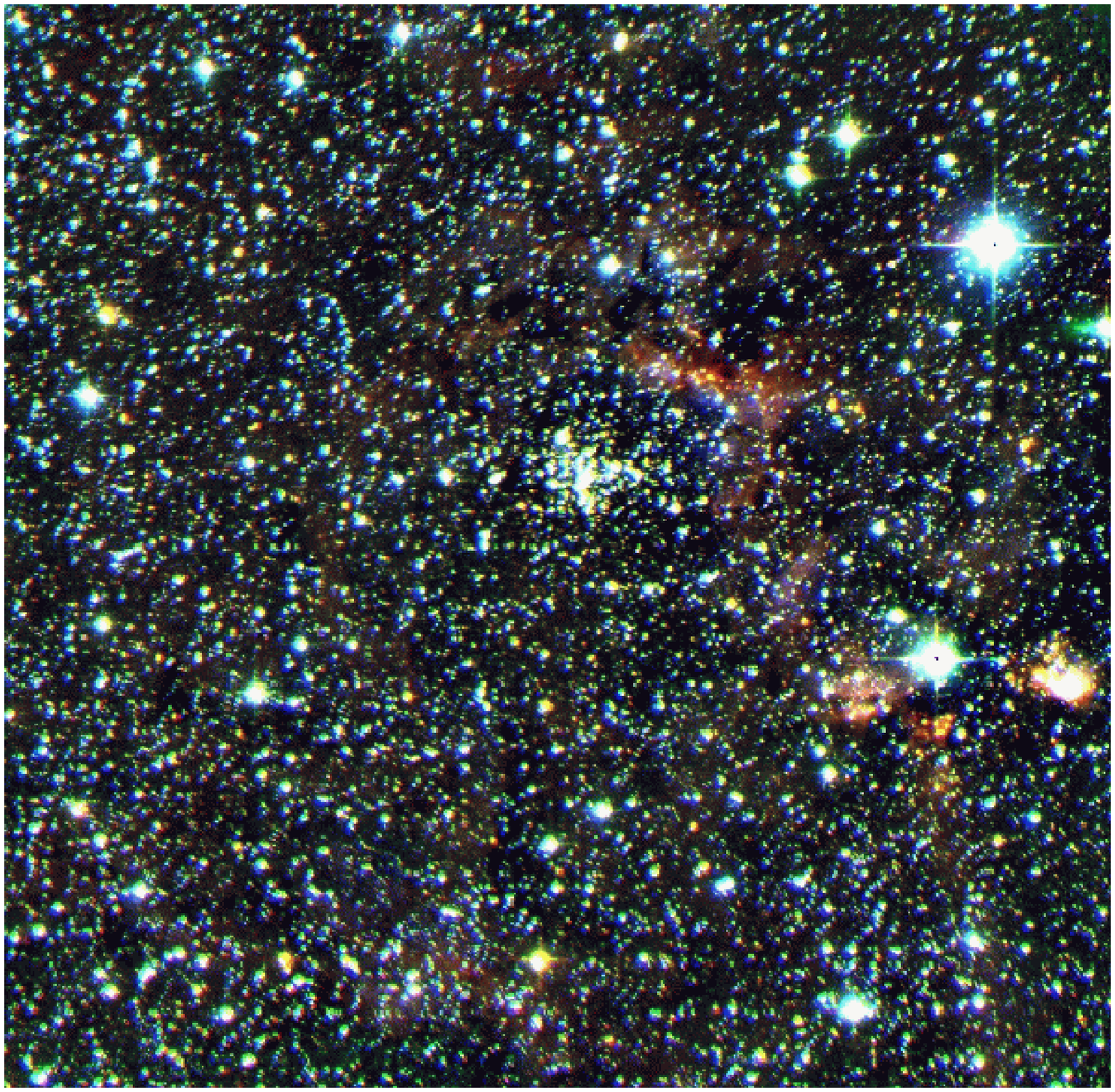}}
\figcaption{\label{afglimage} Color composite ($J$=blue, $H$=green, $K$=red) of the
region around the G305+00.2 cluster.  Image is approximately 8\arcmin\ on a side.  The
cluster is clearly apparent as a concentration of stars with similar colors; no
nebular emission is apparent in the immediate vicinity of the cluster though a
ridge of nebulosity is present to the northwest.}
\end{figure*}

\begin{figure*}
\includegraphics{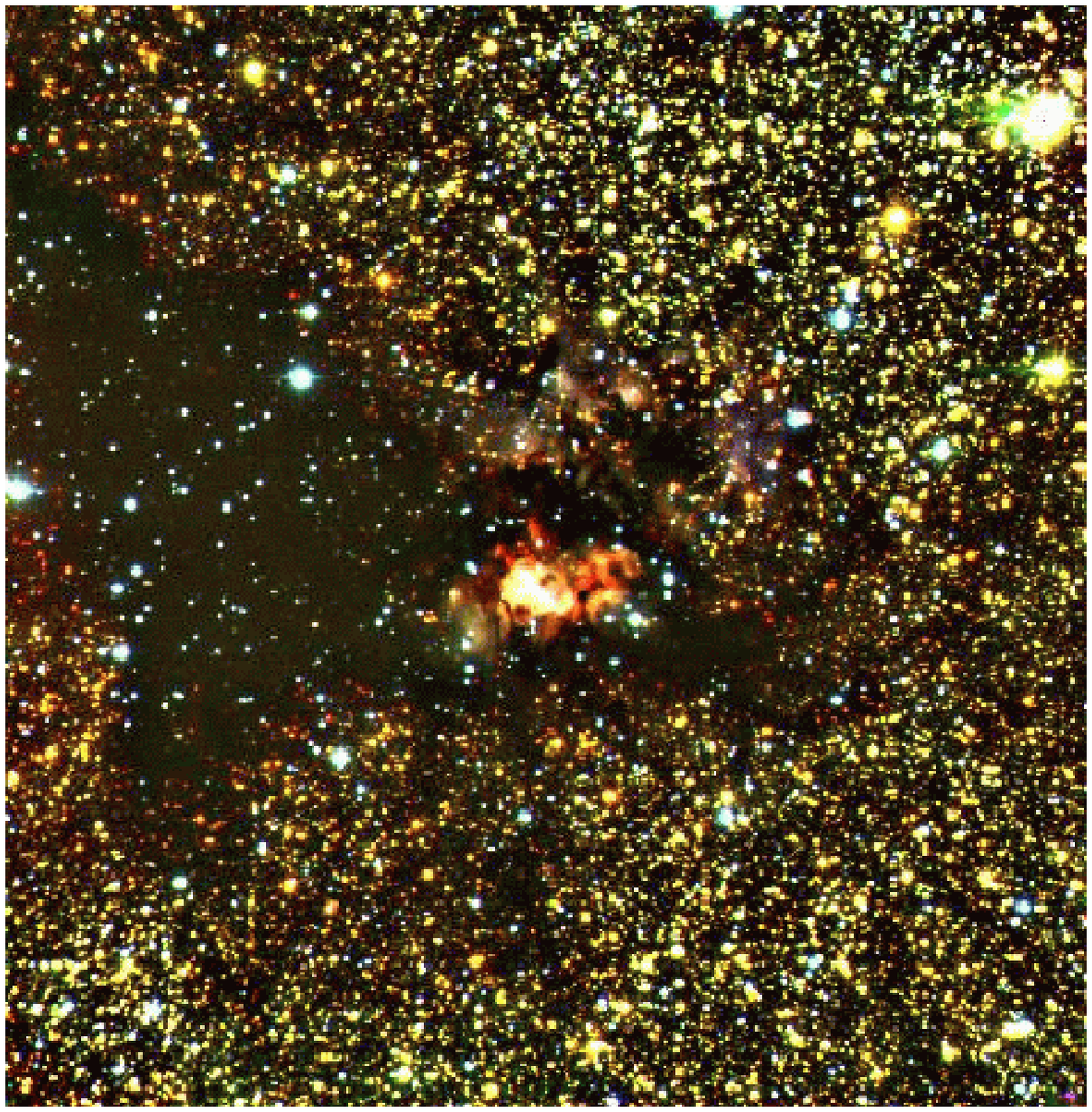}
\figcaption{\label{g353image} Color composite (J=blue, H=green, K=red) of the region around the G353.4-0.36
cluster.  Image is approximately 8\arcmin\ on a side.  The cluster is surrounded by
intense nebular emission and is contained in a larger dark molecular cloud.}
\end{figure*}

We selected a total of four cluster candidates in the southern hemisphere using
the 2MASS Point Source Catalog based on color and density criteria.  Two of the
candidates observed appear to be regions of low extinction and are discussed
elsewhere \citep{cotera05}.  The two confirmed clusters are near radio
\ion{H}{2} regions designated G305.3+00.2 and G353.4-0.4.  We present
three-color composites of the 8\arcmin$\times$8\arcmin\ images of the
G305.3+00.2 and G353.4-0.36 clusters in Figures~\ref{afglimage} and
\ref{g353image} respectively.  

G305.3+00.2 is an \ion{H}{2} region which has
been previously observed using radio recombination lines \citep{wilson70},
\ion{C}{1} emission in the submillimeter \citep{huang}, and in the mid-infrared
(MIR) by the Midcourse Space Experiment (MSX).  The kinematic distance of
$3.5\pm1.1$ kpc obtained for this \ion{H}{2} region \citep{wilson70} agrees
well with the distance of 3.3 kpc for masers several arcminutes away
\citep{g305masers}, suggesting they may be part of a single star-formation
complex.  A distance of 4 kpc is adopted as an upper limit to the radio
kinematic distance by \citet{clarkporter} in a study of the star clusters Danks
1 and 2 in this region.  The situation is more complex for the G353.4-0.36
cluster, which is in a region known to be a site of massive star formation.
There are numerous radio sources located within 1\arcmin\ of the NIR cluster,
which we discuss in detail in \S\ref{g353-discussion}.

All photometric observations were done in excellent seeing conditions:
0\farcs7-0\farcs9.  The images were reduced and combined automatically at the
telescope using the ORAC-DR pipeline. ORAC-DR is a generic data reduction
pipeline created at the Joint Astronomy Centre in Hawaii, originally for use
with various UKIRT and JCMT instruments.  Subsequent reprocessing did not
noticeably improve the images, therefore the pipeline processed data has been
used throughout. Source detection, PSF fitting, and photometry was carried out
using IRAF-DAOPHOT, and is discussed in detail in \S\ref{phot}.

All spectra were obtained with a 1\arcsec$\times$ 7\farcm7 slit.  The long-slit
format combined with the high stellar density within the FOV resulted in the
simultaneous observation of multiple stars.  Total integration times ranged
from 10 minutes to 30 minutes, and were chosen to provide adequate S/N for NIR
spectral classification as described in Hanson, Conti, \& Rieke (1996).  After
the data was flat-fielded, grism curvature was removed using the
FIGARO\footnote{FIGARO is part of the Starlink software package available at
http://star-www.rl.ac.uk/} tasks {\it cdist} and {\it sdist}.  Wavelength
calibration was performed using the the OH$^-$ night sky lines and the FIGARO
task {\it arc}.  The uncertainty in the wavelength calibration fit was
determined to be 2.18~\AA.  The FIGARO task {\it irflux} was used both to
flux-calibrate the spectra and remove the telluric absorption using the G2V
standards HD157017 and HD115496.  Both of the standards had intrinsic Br
$\gamma$ in absorption, with equivalent widths of 5.7~\AA\ for HD157017 and of
5.6~\AA\ for HD115496; in each case, the absorption line was removed by fitting
a line to the continuum in the region of the line in the standard star spectrum
prior to flux calibration.  The individual spectra were obtained by extracting
apertures 4-5 pixels wide from the full spectral array, then performing
background subtraction using apertures of the same width on either side of the
source, separated by 2 pixels (0\farcs9).  We also extracted off-source spectra
in each cluster to characterize any nebular emission.

\section{Analysis}
\label{sec:results}
\subsection{Spectroscopy}
\label{sec:spectra}

The development of NIR spectral atlases of nearby massive stars of known
spectral type (\citealt{hanson96}; \citealt{morris96}; \citealt{blum97}),
provides a valuable classification scheme for stars too heavily obscured by
dust to permit optical spectroscopy.  In the $K$ band, in addition to the
Br$\gamma$ (2.165~\micron) line, massive O stars have helium (\ion{He}{1}\
2.058~\micron, \ion{He}{1}\ 2.112~\micron, \ion{He}{2}\ 2.189~\micron), carbon
(\ion{C}{4}\ 2.078~\micron), and nitrogen (\ion{N}{3}\ 2.116~\micron) lines in
their spectra which allow for the determination of the spectral type to within
a subtype if there is adequate ($\gtrsim 70$) line signal to noise.  Table~6 of
\citet{hanson96} indicates that in many cases the mere presence of these lines
in emission or absorption (without considering equivalent width) is sufficient
to determine spectral type to within two subtypes for O stars.  The situation
is more complicated for B stars, which have fewer features in this part of the
spectrum; however, they are still classifiable using only $K$-band spectra.
 
We obtained $K$-band spectra of five stars in the G305.3+0.2 cluster field and
three stars in the G353.4-0.36 cluster.  In order to reduce the level of
foreground contamination, we imposed a color cut of $H-K > 0.5$ based on the
2MASS magnitudes and selected the brightest stars meeting this requirement.
Despite this cutoff, two of the five stars observed in the G305.3+0.2 cluster
proved to be foreground contaminants with sufficient line-of-sight extinction
to push them over our threshold.  The cluster sequence was much narrower and
more well-separated from the foreground in the G353.4-0.36 cluster, and no
obvious foreground contaminants were present in our spectroscopic sample.  The
G353.4-0.36 cluster was sufficiently red ($H-K_{cluster} \gtrsim 1.3$), that the
time required to obtain a useful signal-to-noise in $H$-band spectra would have
been prohibitively large, so only $K$-band were obtained.

\subsubsection{G305.3+0.2 Cluster}
\label{sec:resafgl}

\begin{figure}
\includegraphics[angle=0,width=9cm]{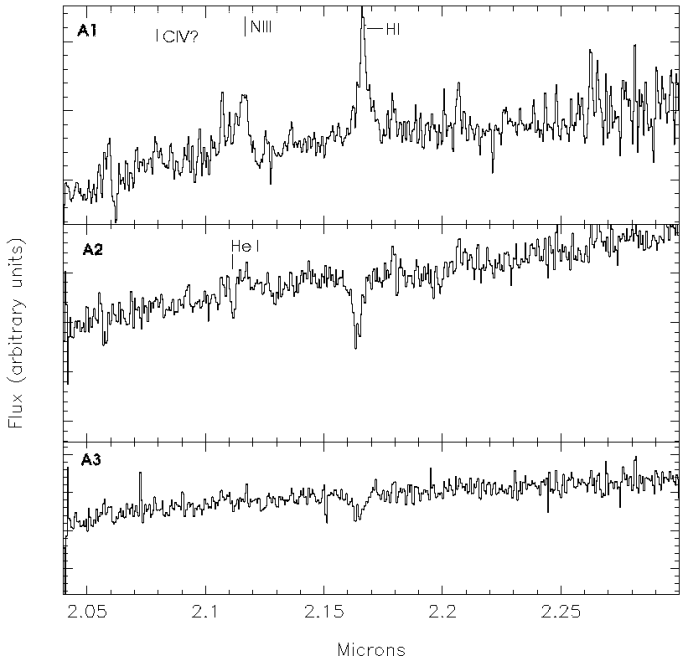}
\figcaption{\label{afglspectra} Spectra for the cluster stars in the G305.3+0.2 cluster.  Top panel:
Source A1, identified as O5-6V.  Middle panel: Source A2, identified as B0-1V.  Bottom
panel: Source A3, identified as B2V-B3V.}
\end{figure}

We present spectra for the three cluster members, which we label A1--A3, in
Figure~\ref{afglspectra}.  

\begin{figure}
\includegraphics[angle=0,width=9cm]{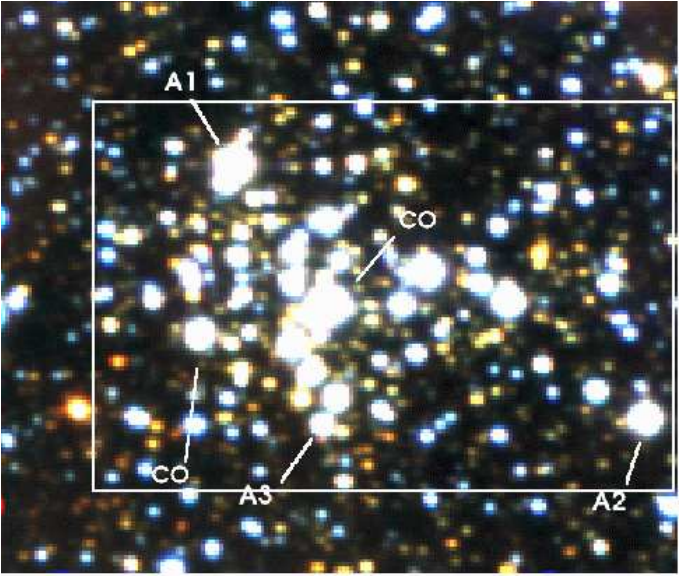}
\figcaption{\label{afglzoom} Region immediately surrounding the G305.3+0.2
  cluster.  Spectroscopically classified sources are marked as A1, A2, and A3
  and sources showing CO absorption are labeled with ``CO''.  This image is approximately 120\arcsec x 106 \arcsec and the ``cluster'' area is marked.}
\end{figure}

In Figure~\ref{afglzoom} we present a 106\arcsec
$\times$ 120\arcsec\ image of the cluster and label the positions of sources
A1--A3.  

The measured magnitudes (see \S\ref{phot}) and observed spectral lines
for A1--A3 are presented in Table~\ref{AFGLews}.  The other two stars for which
we obtained high S/N spectra have late-type spectra, as indicated by strong CO
absorption at 2.29 and 2.32~\micron, suggesting they are either foreground
objects or YSOs.  The lack of nebular emission in the cluster and the presence
of weak (nearly the same as in the G2V spectral standard) Br $\gamma$
absorption in one of the spectra suggest that these are foreground objects
rather than YSOs.  In addition, the $K$ magnitudes of these objects ($K = 9.43$
and $K = 10.114$) make them too bright to be low-mass YSOs at the cluster
distance, and the presence of main-sequence O and B stars argues against
identifying these objects as massive YSOs.  We thus conclude that these two stars are most
likely late-type foreground stars, and excluded them from further analysis.

\begin{figure}
\includegraphics[angle=0,width=9cm]{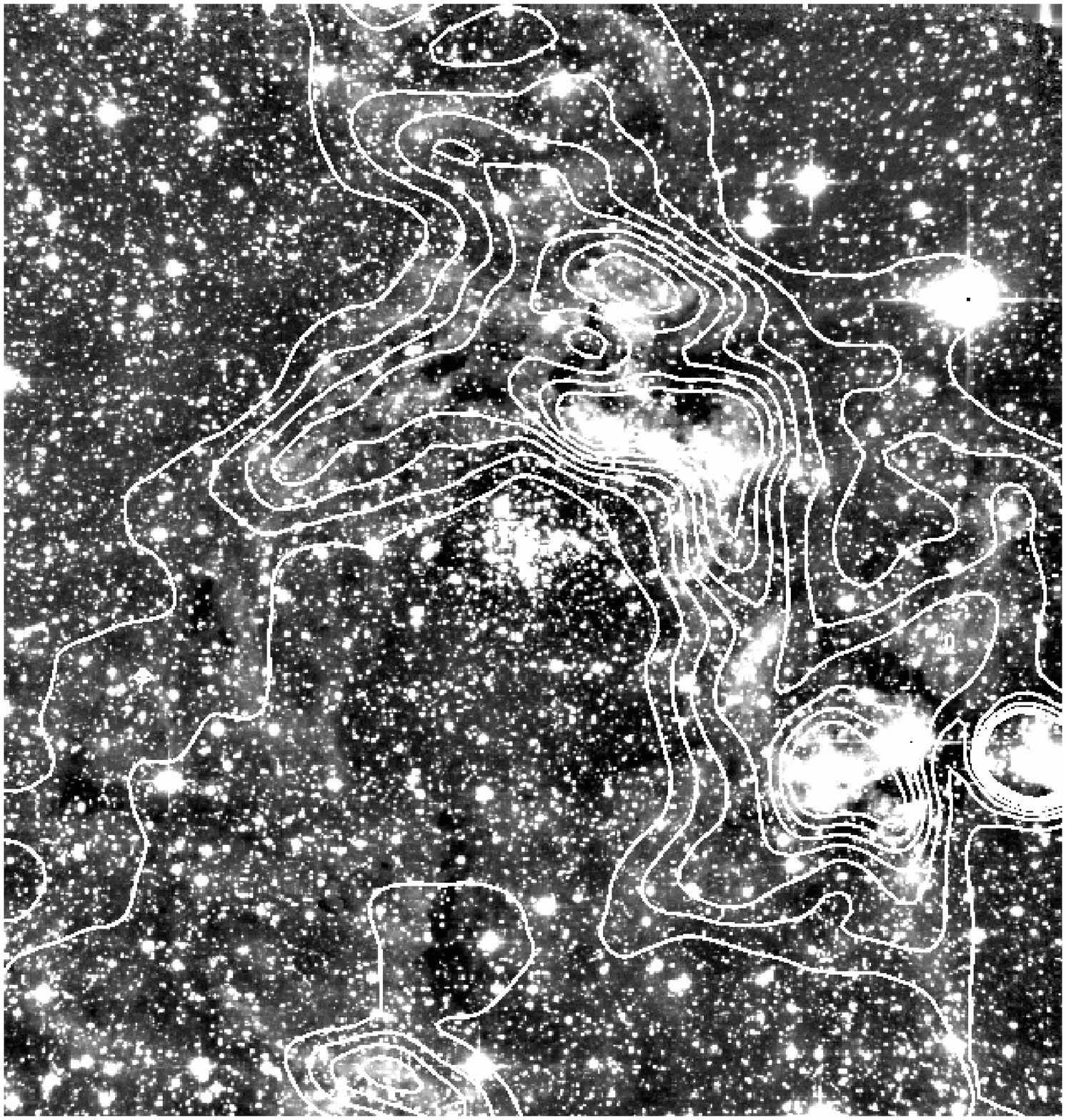}
\figcaption{\label{msximage} $K$-band image of the G305.3+0.2 cluster region with 8~\micron\ contours
  from the MSX mission.  The $K$ image has been stretched to emphasize the
  nebular emission.  Note the close correspondence between the mid-IR
  emission and the nebular $K$-band emission.}
\end{figure}

Although nebular emission can be seen in the full image
(Figure~\ref{afglimage}), it is significantly removed ($\gtrsim$1\arcmin) from
the cluster. Nevertheless, in order to ensure that any measured Br$\gamma$
(2.166~\micron) is stellar in origin and not contaminated by nebular emission
within the cluster, we extracted a local background spectrum for the cluster.
There were no features apparent in the resulting spectrum; we thus conclude
that nebular emission within the cluster is negligible.  This conclusion is
supported by an apparent bubble of MIR emission seen in the MSX Band A image
(see Figure~\ref{msximage}); the MIR emission avoids the cluster itself.

Figure~\ref{afglspectra} shows that source A1 has emission lines with
equivalent widths stronger than -2~\AA\ at 2.116~\micron\ and 2.166 \micron\
(see Table~\ref{AFGLews}).  The line at 2.166 \micron\ is immediately
identifiable as Br$\gamma$.  We identify the line at 2.116\micron\ as
\ion{N}{3}, which is consistent with the lines used in the the classification
system presented in \citet{hanson96}; the broad nature of this line is due to
the multiplet nature of the transition responsible rather than broadening by
stellar winds.  The presence of Br$\gamma$ and \ion{N}{3} 2.116\micron\ in
emission, without further information and without equivalent widths, is
sufficient to identify the star as being an early to middle O supergiant; the
broad Br$\gamma$, produced in the stellar winds, is not observed in
main-sequence O stars \citep{hanson96}.  There is a possible weak detection
($\sim 2\sigma$) of \ion{C}{4} in emission at 2.078~\micron.  This line only
appears in O stars ranging from O5 to O6.5 \citep{hanson96}, and if real,
significantly constrains the stellar type.  Helium lines are often observed
both in emission and absorption in the spectra of massive stars: \ion{He}{1}\
(2.058~\micron), \ion{He}{1}\ (2.112~\micron), and \ion{He}{2}\
(2.189~\micron), are all absent from the spectrum of A1.  Poor removal of the
telluric features near the 2.058~\micron feature prevents us from drawing any
conclusions based on our non-detection.  If real, the absence of the
\ion{He}{1}\ (2.112~\micron) line restricts the spectral type to O6 or
earlier. A \ion{He}{2}\ line is expected in an O star; by
estimating the strength of possible features dominated by the noise (as
described in detail in \S~\ref{sec:g353}) we can place an upper limit of 0.5
\AA\ on the equivalent width of any potential \ion{He}{2}\ (2.188~\micron)
feature.  This is consistent with the width of the feature in the stars
observed by \citet{hanson96}, so the non-detection does not rule out an O star
identification for this source.  Taken together, these spectral characteristics
suggest a spectral type of O5Ib-O6Ib for Source A1.  If the weak detection of
\ion{C}{4} is discounted, the presence of the \ion{N}{3}\ line and the limit on
an \ion{He}{2} line at 2.188 \micron\ allows an O7-O8 identification as well.
Even when present, however, the \ion{C}{4}\ line is weak, with an equivalent
width weaker than -2~\AA; thus, while a positive detection of this line would
allow for definitive classification of this source as an O5Ib-O6Ib star, a
non-detection at the given S/N does not preclude the same classification.

The intrinsic NIR colors of O and B stars range from -0.08 to -0.01
\citep{wegner}; this small range allows an extinction to be derived even
without knowing the precise spectral type of a massive star.  For source A1,
the extinction thus derived based on the observed $H-K$ color is $A_V = 12$
assuming the extinction law of \citet{extinction}.  However, the large range in
absolute $M_K$ for O supergiants prevents us from making a distance
determination based on Source A1.  We can only say the distance is greater than
$\sim 3.3$ kpc, which would be the distance for a main-sequence O5-O6 star.
\citet{clarkporter} adopt a distance of 4 kpc to the Danks 1 and 2 clusters in
the same star formation complex, calling it an upper limit to the values
allowed by the radio and H$\alpha$ observations, and we will follow suit,
acknowledging that the uncertainties in this value are $\sim 0.5$ kpc.

Source A2 shows a strong Br$\gamma$ (2.166~\micron) line in absorption with an
equivalent width of $6.2 \pm 1.2$~\AA\ and a probable weak \ion{He}{1}\ (2.112
\micron) line in absorption with EW = $0.7\pm 0.2$~\AA.  This combination of
features occurs only in B stars; a comparison of the equivalent width of the
lines with the B stars of \citet{hanson96} suggests an spectral type in the
range of B2-B4.  If the \ion{He}{1}\ line is considered only as an upper
limit, the classification becomes more problematic, and the star could range
from B2-A2.  The star has $H-K = 0.68$, which for any star in this range of
spectral type excludes a foreground object.  Unlike for A1, the luminosity
class of these sources cannot be determined from these spectral features; as
\citet{hanson96} points out, the $K$-band spectra of early B supergiants are 
indistinguishable from those of early B main-sequence stars about half the
time, and those of late-B supergiants cannot be distinguished from early-B
dwarfs. 

If we assume that A2 is a cluster star, we can constrain the absolute
magnitude, and thus the spectral type, by requiring the distance to be the same
as for the O star. Since the intrinsic near-infrared colors vary by less than
0.1 magnitude for stars in the range of spectral types allowed by the spectrum
\citep{wegner}, we can derive a extinction for this source rather than use that
derived from the O star, thus reducing the effects of differential extinction.
This gives an extinction to source A2 of $A_V = 11.6$, or $A_K = 1.3$ using the
reddening law of \citet{extinction}.  At the distance of $4$ kpc, we obtain an
absolute magnitude for Source A2 of $M_K = -4.0$, roughly that expected for an
O8V star.  This identification is not consistent with the spectral features of
A2; a smaller distance, or an identification of A2 as an early-B supergiant,
could explain the spectrum of A2.  If the radio distance of $3.3 \pm 0.3$ kpc
is used instead, we obtain an absolute magnitude of $M_K = -3.3$ for Source A2,
making it a B0V-B1V.
 
Source A3 shows only Br$\gamma$ in absorption with an EW of $5.9 \pm 1.3$
\AA. We place an upper limit on an \ion{He}{1} absorption line at 2.112
\micron\ of 0.6~\AA.  As discussed above, this width for Br$\gamma$ only
constrain the classification of the star as main sequence B or early A.  The
observed $K$ magnitude is $11.96$, which corresponds to an absolute $M_K \simeq
-2.6$ assuming the extinction and distance of an O5Ib-O6Ib star for source A1;
this is consistent with an identification of A3 as a main-sequence B1V star.
The radio distance would imply a B2V identification, also consistent with the
spectral features of A3.  Source A3 is not among the brightest stars in the
cluster region; it happened to fall in the same long slit as one of the
foreground contaminants we had targeted for observation.  This suggests that
the other cluster members brighter than A3 are also late O or early B stars.

\subsubsection{G353.4-0.36 Cluster}
\label{sec:g353}

\begin{figure}
\includegraphics[angle=0,width=9cm]{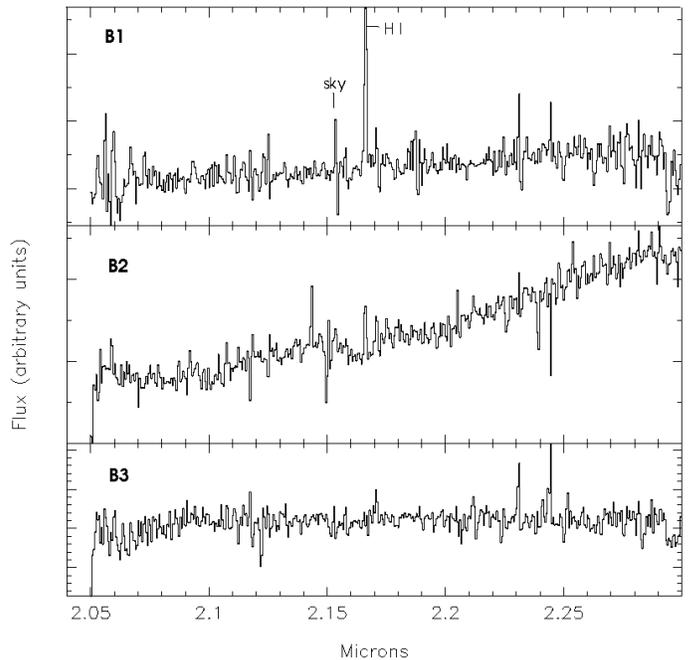}
\figcaption{\label{g353spectra} Spectra for sources in the G353.4-0.36 cluster.  All three are identified as 
massive YSO candidates.  The Br$\gamma$ emission line seen in B1 is
contaminated by nebular emission (see Fig.~\ref{G353_offsource}).}
\end{figure}

Spectra for the three sources observed in this cluster are presented in
Fig.~\ref{g353spectra}.

\begin{figure}
\includegraphics[angle=0,width=9cm]{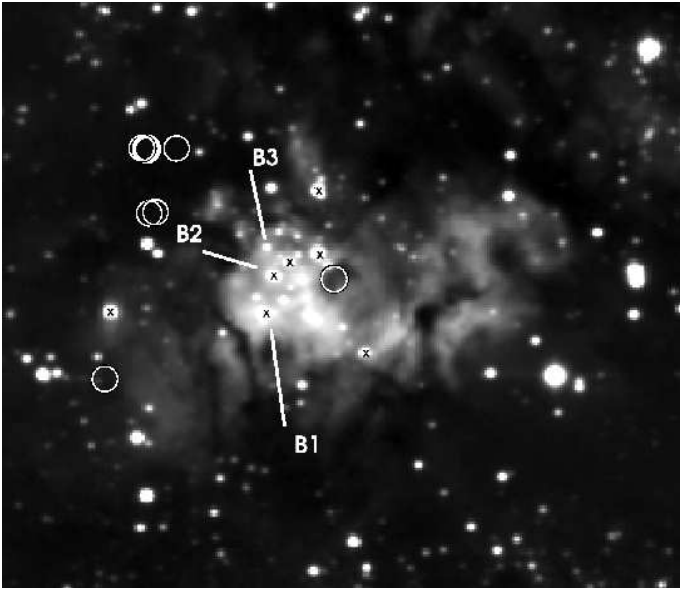}
\figcaption{\label{ginset} Maser positions from the literature (\citet{masers,masers2,masers3})
  overlaid on the G353.4-0.36 cluster K-band image.  Note that they appear in
  regions which are dark in the near-IR, suggesting a more deeply embedded
  origin. Sources B1-B3 are indicated, and all cluster sources detected in $H$
  and $K$ are marked with crosses.}
\end{figure}

An enlarged version of the relevant portion of
Fig.~\ref{g353image} is presented in Fig.~\ref{ginset}, with the positions of
the spectroscopic targets indicated with arrows and labels.  

\begin{figure}
\includegraphics[angle=0,width=9cm]{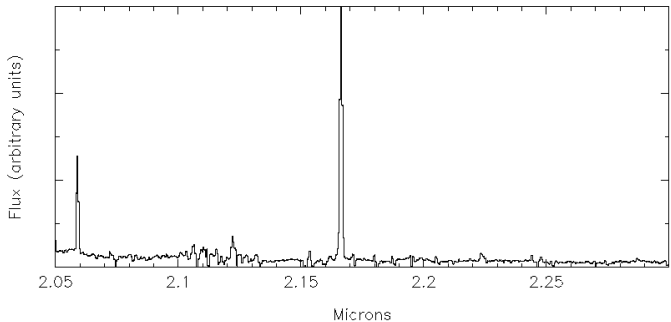}
\figcaption{\label{G353_offsource} Nebular spectrum from the G353.4-0.36 cluster region.  Emission lines present are \ion{He}{1} 2.058~\micron, H$_2$ 2.12~\micron, and Br$\gamma$ 2.166~\micron.}
\end{figure}

The only
non-nebular feature which we detect is CO absorption in Source B1; the Br
$\gamma$ emission observed in all three spectra is contaminated by nebular
emission to such a degree that we cannot disentangle any stellar component that
may be present.  While this line is much stronger in B1 than in the other two
sources, the nebular emission is highly spatially variable in the cluster
region and this does not demonstrate a stellar origin for the line.
Additionally, the line width is significantly narrower than that of Source A1
and is similar to that observed in the off-source nebular spectrum
(Fig.~\ref{G353_offsource}).  

The CO absorption in Source B1 in combination
with the red colors (Table~\ref{g353table}) are similar to those associated
with solar-mass young stellar objects (YSOs) \citep{greenelada96}, or a cool
giant or supergiant.  If B1 is a YSO, the CO absorption is from the
circumstellar material; otherwise it is photospheric in nature.  Using the
radio kinematic distance \citep{forster00} of 3.6 kpc to the cluster, we derive
an $M_K$ for Source B1 of -0.8 without correcting for extinction.  Correcting
for extinction is difficult to do accurately in this region of highly variable
extinction, especially when the intrinsic colors are not known since the nature
of the object is uncertain.  Nevertheless, limits can be placed on the amount
of extinction present, and thus the absolute magnitude of Source B1.  The lower
limit is given by the uncorrected value of $M_K = -0.8$, which assumes the
color observed is the intrinsic color, while the bright limit can be derived
assuming an intrinsic $H-K = 0.3$, characteristic of late-type stars; this
gives an extinction to source B1 of $A_V = 16.6$ magnitudes and an
extinction-corrected absolute $M_K$ of $-2.6$.  This is several magnitudes
brighter than the expected magnitude of YSOs of approximately a solar mass at
the distance and extinction of this cluster, $M_K \sim 1-3$ \citep{oasa}, and
somewhat lower than the $M_K$ for massive YSOs, $M_K \sim -1$ to $-5$
\citep{ishii}.  Finally, we note that this $M_K$ is consistent with that for a
$7 M_{\odot}$ YSO \citep{chakrabotry}.  We conclude that if Source B1 is a YSO,
it has a mass greater than a few solar masses based on its absolute magnitude
in $K$, but observations of more massive YSOs are still sufficiently few that a
more accurate mass determination based solely on the absolute magnitude is not
possible.  Given the nebular emission, seen as \ion{He}{1} (2.058~\micron),
H$_2$ (2.12~\micron), and Br$\gamma$ (2.166 \micron) emission off the stellar
sources (see Figure~\ref{G353_offsource}), G353.4-0.36 is obviously a region of
current star formation; therefore, the identification as a massive YSO is more
probable than a late type cool giant or supergiant located in the cluster
itself.

Since Source B1 was not detected in $J$, it cannot be placed on a color-color
diagram to determine whether a NIR excess is present, which could help to
discriminate between the YSO and cool field star possibilities.  For B1 to be a
cool giant, it would need to be a foreground star with the appropriate color
and magnitude, which falls by chance in the cluster region.  Rather than use
the entire 8\arcmin $\times$ 8\arcmin\ field to determine the field star
density, as we did for the G305.3+0.2 cluster (\S\ref{phot}), we used
only the heavily extincted region surrounding the cluster.  This is because the
molecular cloud in which the cluster is embedded extinguishes the background
stars to such a degree that using the entire field would significantly
overestimate the level of field star contamination in the immediate region of
the cluster.  We estimate the probability of a field source as bright as or
brighter than Source B1 and red enough to satisfy the color cut falling within
the cluster region to be approximately 18\%.  This is a conservative estimate,
since at the edges of the cloud reddened sources become visible and increase
the field star density, especially of red objects, over what it would be at the
location of the cluster.  Nevertheless, we cannot rule out either a foreground
giant or a YSO explanation for Source B1.

As with Source B1, the non-detection of Sources B2 and B3 in $J$ prevents us
from using a color-color diagram to measure NIR excess.  No photospheric
features are detected in the spectra of either Source B2 or B3; Source B2 shows
a rising spectrum in $K$ suggesting a strong NIR excess, while the spectrum of
B3 is essentially flat in this region.  In order to determine whether the
spectra were truly featureless or merely had a signal-to-noise too low to see
expected features, we fit a continuum to the spectra and examined all
excursions above and below the fit.  90\% of these deviations had an equivalent
width less than 1.7~\AA.  For comparison, the detected absorption lines
tabulated by \citet{greenelada96} for low-mass YSOs range in equivalent width
from 0.3-5.6~\AA\ for Na I and Ca I, with CO usually exceeding 2~\AA\ when
present.  \citet{ishii} conducted a similar survey of massive YSOs; the only
emission lines other than Br $\gamma$ detected in a significant number of
sources are CO (with an equivalent width exceeding 4~\AA) and H$_2$ (with an EW
$> 3$~\AA\ in all cases, and $> 5$~\AA\ in most cases).  We thus conclude that
Source B2 is genuinely featureless, but cannot classify it.  The final source,
Source B3, had no reliably detected features but the signal-to-noise was low
enough that we cannot reliably call it featureless.

The observed $K$ magnitudes are consistent with a B star identification for
sources B2 and B3; however, the extincted but distance-corrected $M_K$
magnitudes of $\simeq -0.2$ to $-0.6$ are also similar to those observed for
the massive YSO ($M \simeq 7 M_{\odot}$) 05361+3539 \citep{chakrabotry}.  Thus,
although these sources are massive, we cannot distinguish based on their NIR
spectra or magnitudes between shrouded B stars and less-evolved YSOs.  Mid-IR
observations with sufficient resolution to resolve the individual sources
(separated by $\sim 5$\arcsec) would aid in this determination; deeper J-band
photometry, detecting more of the cluster stars, would also be useful.  We note
that although we see ionized gas suggesting the presence of O stars, we have
not detected any O stars which would be the source of the ionizing radiation in
this cluster.

Due to the young age of the sources observed in this cluster and the lack of
photospheric features in their spectra, the spectra were unsuitable for
determining a reliable distance.  Thus, the kinematic distance to the
associated maser and UCHII \citep{forster00} was used instead, adjusted to a
distance to the Galactic Center of 8 kpc from the original 10 kpc.  This gave a
distance to the cluster of 3.6 kpc.  Assuming an intrinsic $H-K = 0$, we
estimate the reddening to the cluster at $A_V = 22$ based on the narrow cluster
sequence at $H-K \simeq 1.3$ and assuming the extinction law of
\citet{extinction}.  This estimate is highly uncertain due to the young age of
the sources; many are likely to have a near-infrared excess leading to an
overestimate of the line-of-sight extinction to the cluster.
 
\subsection{Photometry \label{phot}}
\label{sec:images}

We obtained images in $J$, $H$, and $K_s$ of both clusters to a limiting
magnitude of approximately $J=16$, $H=18$, $K_s = 18.5$, with total integration
times of 12 minutes in each band.  The limiting magnitude was brighter than
expected due to confusion, which is most noticeable in $J$ due to the slightly
larger PSF and the greater sensitivity of the instrument at shorter
wavelengths.  Seeing was 0\farcs7--0\farcs8, which, since the IRIS2 platescale
is 0\farcs45/pixel, resulted in a slight undersampling of the point spread
function (PSF), thus making PSF fitting more uncertain.  Our individual images
were taken using a random dither pattern with sub-pixel dithers employed to
improve the PSF.  In an effort to better understand our errors we performed
both PSF fitting and aperture photometry for each source.  There was no
systematic offset between the two methods, but the errors were $\sim 2$ times
larger for the aperture photometry due to the crowded fields.

Photometric calibration was performed using the 2MASS magnitudes of field
stars, after correcting from the IRIS2 filter system to the 2MASS filter system
as described in \citet{carponline}.  The calibrated magnitudes for the stars in
the cluster area are presented in Table~\ref{afgl-phottable}.  The large field
of view and location in the Galactic Plane provided over 100 stars in each
pointing which were bright enough to have good photometry with 2MASS, but faint
enough to be unsaturated in our IRIS2 images $(11.5 < K_s < 14)$.  Those stars
which were relatively isolated in the IRIS2 images were used as the photometric
calibration set.  We chose to use a relatively large number of calibration
stars rather than selecting the few most isolated stars to reduce effects of
potential variability and photometric outliers among the calibration stars.
The scatter in the photometric calibration derived from comparison to 2MASS is
the dominant source of photometric error, contributing two to three times the
measurement errors as reported by DAOPHOT.  DAOPHOT errors were $\simeq 0.03$
mag while the calibration uncertainties were ($\Delta J = \pm 0.05, \Delta H =
\pm 0.06$, and $\Delta K = \pm 0.06$ mag).  Quoted errors in the 2MASS
photometry were negligible, with most stars having an error of $\pm 0.003$ mag
or less in all bands.  Thus, the quoted error should be considered an
overestimate when considering the \emph{relative} photometry of stars within
either cluster; the calibration errors from comparison to the 2MASS photometry
will shift all our measurements by the same amount.  No trend in the
photometric errors, either internally or relative to the 2MASS data, was
observed with location.  Finally, the positions of the stars were also adjusted
to agree with 2MASS by minimizing the offsets between the 2MASS and IRIS2
positions allowing for pointing offset and rotation.

\subsubsection{G305.3+0.2}
\label{g305phot}
The color composite of the full $J$, $H$, and $K_s$ images is presented in
Fig.~\ref{afglimage}; the cluster alone is shown in Fig.~\ref{afglzoom}, with
the spectroscopic targets marked.

The cluster is clearly visible in the
full-size image with a concentration of nebular emission to the northwest.  In
order to help determine whether the nebular emission is physically associated
with the cluster, we overplotted the contours at 8~\micron\ from the MSX
mission\footnote{On-line data are available from
http://www.ipac.caltech.edu/ipac/msx/msx.html}.  (Fig.~\ref{msximage}).  The
ridge of near-IR nebulosity corresponds to the brightest portion of a roughly
circular structure of mid-IR emission, with the cluster located in the interior
where there is no mid-IR emission present.  The general appearance is that of a
wind-blown bubble, and the 8~\micron\ emission wraps entirely around the
cluster at a lower level.  The cluster is located off-center in this structure,
near the brightest portion of the mid-IR emission, but there is no mid-IR
emission and no near-IR nebulosity present in the area of the cluster
itself. The cluster is dense and well-defined, with stellar density much higher
than in the field.

\begin{figure}
\includegraphics[angle=0,width=9cm]{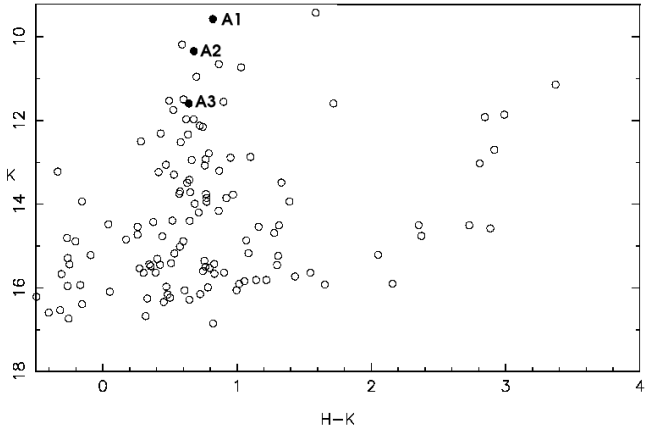}
\figcaption{\label{afglcm} K vs H-K for the G305.3+0.2 cluster region, with all sources
plotted. Typical error bars are smaller than the circles.  Filled symbols
designate spectroscopic targets A1-A3.}
\end{figure}

The $K$ versus $H-K$ color-magnitude diagram of the cluster region is shown in
Figure~\ref{afglcm}.  

\begin{figure}
\includegraphics[angle=0,width=9cm]{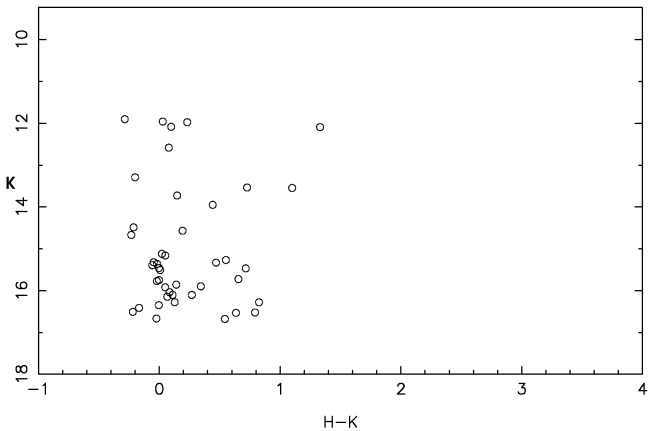}
\figcaption{\label{afgl_control} K vs H-K for a randomly selected control field for G305.3+0.2, with the same area as
  the cluster field.}
\end{figure}

At radii of approximately 30\arcsec\ in the east-west
direction and 20\arcsec\ in the north-south direction from the cluster center
the stellar density has fallen to that of the field, which we used to define
the cluster region.  Foreground stars are apparent in the color-magnitude
diagram at $H-K \sim 0.3$; in this cluster there is no clear separation in
color between cluster and field stars, just an overdensity of redder stars in
the cluster; as a result, we cannot impose a firm color cut to separate field
stars from cluster stars.  A color-magnitude diagram of a randomly selected
control field with the same area as the cluster is shown in
Fig.~\ref{afgl_control}; many fewer stars are present, especially at bright
magnitudes and moderately red colors.

\begin{figure}
\includegraphics[angle=0,width=9cm]{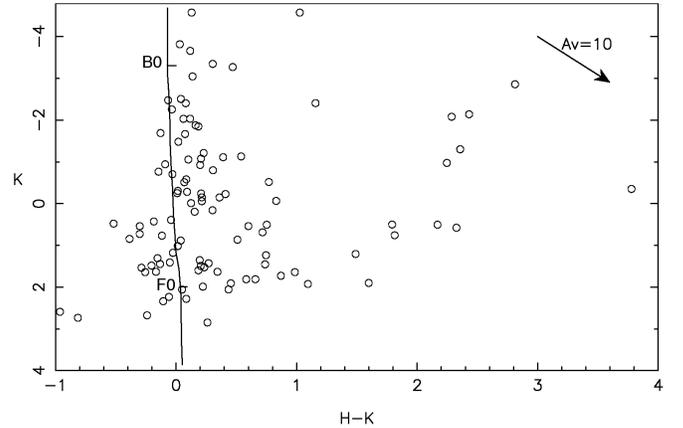}
\figcaption{\label{ZAMSa}
Distance- and reddening-adjusted K vs. H-K for the G305.3+0.2 cluster
region, statistically corrected for field star contamination.  The ZAMS from the \citet{newmodels} evolutionary models has been
transformed to the observed quantities and overplotted.}
\end{figure}

In order to account for field star contamination within the cluster region, we
determined the average number of stars per square arcminute in the image
outside the cluster region in color-magnitude bins of $\Delta K = 0.5$, $\Delta
(H-K) = 0.5$ and randomly selected the appropriate number of stars from the
cluster field for removal.  This is similar to the procedure employed by, among
others, Blum, Conti, \& Damineli (2000) and \citet{figueredo02}.  In cases
where less than one star was expected in the cluster field in a particular
color-magnitude bin, the number expected was used as a probability for removing
a star.  A total of 24 ``field'' stars were removed, leaving 115.  The main
concentration of cluster stars is at about $H-K = 0.8$, with a gradually
declining number present out to $H-K \sim 4$.  The resulting cluster CMD with
the field stars statistically removed is shown in Fig.~\ref{ZAMSa}.

Given the
spectroscopically confirmed presence of OB stars in the cluster, as well as the
lack of an obvious color gap, we consider it more likely that these very red
sources are either background sources or sources with a near-IR excess due to
local dust than that they represent a separate cluster giant branch.  The red
sources are not concentrated toward any part of the cluster, though they may
occur more frequently on the outskirts (as would be expected if they are
background objects).  Sources redder than $H-K = 1.5$ were excluded from
analysis of the cluster KLF and IMF; they are unlikely to be main-sequence
cluster members.  If they are included and assumed to be on the main sequence,
the resulting extinction correction would give very large values for the masses
and an overly flat slope to the IMF.  If these sources are cluster members,
they are pre-main-sequence objects, and their masses are difficult to determine
from $H$ and $K$ photometry alone.  Thus, including them in the IMF
determination would give an inaccurate result whether or not they are cluster
members, and they have been excluded.  Finally, the crowded nature of the
cluster region means that these very red sources may suffer from poor
photometry.  

\begin{figure}
\includegraphics[angle=0,width=9cm]{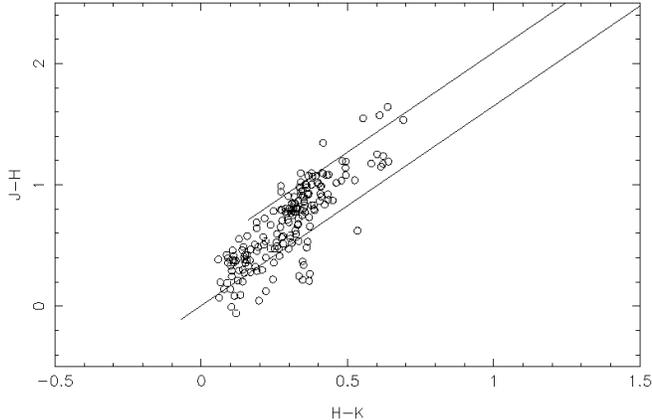}
\figcaption{\label{g305colcol}
Distance- and reddening-adjusted $J-H$ vs. $H-K$ for the G305.3+0.2 cluster
region, statistically corrected for field star contamination.  Error bars are
comparable to the size of the points or smaller.  Very few sources
fall outside the region of reddened main-sequence stars by more than 2$\sigma$.}
\end{figure}

The $J-H$ vs. $H-K$ color-color diagram (Fig.~\ref{g305colcol}) is of limited
utility in identifying cluster members or determining whether some cluster
members are pre-main-sequence objects.  

Since many sources were undetected in
$J$, it will not represent all cluster members, and faint red sources (where we
would expect to find the relatively low-mass, pre-main-sequence objects) would
be most commonly missed in the color-color diagram.  A cut based solely on
$H-K$ must still be applied to exclude background sources.
Fig.~\ref{g305colcol} shows few sources in the area occupied by
pre-main-sequence objects.  Of those sources separated from reddened
main-sequence stars by more than 3$\sigma$, three are relatively faint sources adjacent to
bright sources and one is in a particularly crowded region.  The remaining
three could potentially be pre-main-sequence objects.  However, due to the lack of
observed gaseous emission from the cluster, we consider it unlikely that these
are truly pre-main-sequence stars, and exclude them from the analysis along
with the objects in the unphysical blue region of the color-color diagram
as likely suffering from blending or a mismatch between sources in the
different bandpasses.  There are few enough sources in this region that we do
not expect their inclusion or exclusion to greatly affect the IMF
determination.  

\subsubsection{G353.4-0.36}
\label{g353-discussion}

The $J$, $H$, and $K$ color composite of G353.4-0.36 is presented in
Fig.~\ref{g353image}.  The youth of this cluster is immediately apparent from
its heavily embedded nature and the dense molecular cloud that surrounds it.
This region has long been known to be a site of massive star formation, and it
has been studied extensively in the radio and sub-mm, including continuum
observations at 1.5 GHz, 5 GHz \citep{becker}, and 850~\micron\
\citep{carey2000} as well as molecular line observations in CS
\citep{whiteoakcs}, CO \citep{whiteoakco}, H$_2$CO \citep{whiteoakh2co}, HNCO
\citep{zinchenko} (identified as a dense molecular core), and SiO \citep{sio}.
These signatures of ongoing star formation, combined with the strong nebular
emission still present around the sources observed spectroscopically, suggest
that the cluster is quite young, without main-sequence stars.  Many of the
continuum and molecular line observations quote slightly different positions
for the source peak, and sources separated by several tens of arcseconds are
all identified with the IRAS point source 17271-3439.  Since the beam sizes in
many instances are comparable to the size of the NIR-bright nebulosity and to
the separation between sources, it is likely that the extended source
measurements are observing the same complex, which may peak at different
locations in different wavelengths.  Many of the radio data are tabulated by
\citet{chan96}, who identify a massive YSO in the region based on the IRAS
colors.  It is obvious from the NIR imaging that this source is not a single
point source; in addition to the NIR sources, there are at least four separate
sets of masers \citep[e.g.][]{masers,masers2,masers3}, one of which is
associated with an UCHII \citep{forster00}.  Positions of the masers are
indicated in Fig.~\ref{ginset}.  We note that the masers occur in regions which
are heavily extincted in the near-IR.  OH, H$_2$O, and CH$_3$OH masers are all
known in the region; the latter in particular are indicative of ongoing massive
star formation.  Clearly the sources visible in the near-infrared are only the
tip of the iceberg, with other massive stars still in the process of formation.
Higher-resolution maps at radio and sub-mm wavelengths are necessary to obtain
a full understanding of this region.

In the region of the large dark molecular cloud, only foreground stars are
visible.  This implies $A_V > 50$ in order to completely obscure the stars even
in $K$, assuming a $K$-band detection limit of 17 and a distribution of $K$
magnitudes similar to the rest of the field.  The less heavily extincted region
in which the cluster is visible in the near-IR must have been partially cleared
out by stellar winds and ionization from massive stars.  The relative position
of the NIR stars and the methanol masers (which lie in regions of higher
extinction) suggest that we are observing stars nearer the main sequence which
are emerging from the dust, while objects at an earlier evolutionary state are
offset from this region, indicating ongoing star formation.

\begin{figure}
\includegraphics[angle=0,width=9cm]{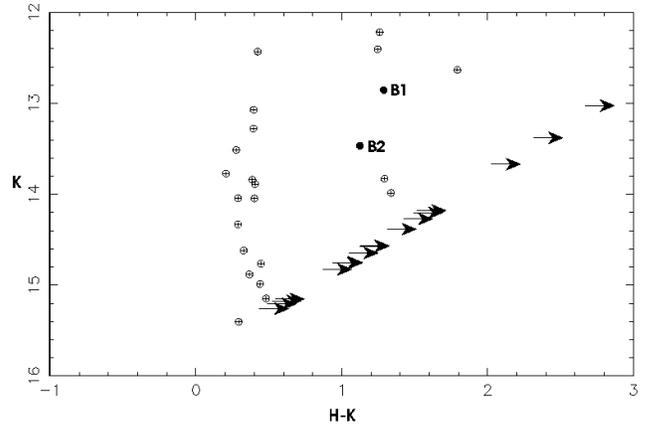}
\figcaption{\label{gcolmag} K vs H-K for the G353.4-0.36 cluster region.  Note the clear separation
in color between cluster and foreground sources.  Sources B1 and B2 are denoted
by filled symbols; Source B3 was not detected in H.}
\end{figure}

The color-magnitude diagram of the G353.4-0.36 cluster is presented in
Figure~\ref{gcolmag}.  

The cluster sequence is much narrower and more
well-separated than in the G305.3+0.2 cluster, allowing for reliable separation
of foreground objects based solely on $H-K$.  Thus, we did not carry out a
statistical removal of foreground objects for this cluster, instead considering
only the objects well-separated from the foreground sequence.  Due to the high
extinction toward this cluster, a large number of objects in the cluster area
were detected only in $K$ (shown as limits in Figure~\ref{gcolmag}).  The KLF is
thus likely to be more reliable than the color-magnitude and color-color
diagrams in determining cluster characteristics.

\section{The K Band Luminosity Function and the Initial Mass Function}
\label{sec:klfimf}

Once field stars have been rejected as described in \S\ref{g305phot}, we can
compute the KLF for both clusters.  For the G305.3+0.2 cluster, which has more
than 100 stars remaining, we additionally compute the initial mass function
(IMF) using two different techniques, the first using the KLF and the second
using the color-magnitude diagram and the spectroscopy of the massive stars.
The KLF is commonly used to determine the IMF even when multi-color photometry
is available; we take this opportunity to test the robustness of this method
and compare the results between this simple and commonly used method and the
more involved method using the color-magnitude diagram.  This will help to
understand the uncertainties and systematic errors that may be a factor when
only the KLF method can be used to derive an IMF.  There were too few
stars to robustly compute the IMF for the G353-0.4 cluster, so we compute only
the KLF in this case.

\subsection{The G305.3+0.2 Cluster}
\label{afglimf}

\begin{figure}
\includegraphics[angle=0,width=9cm]{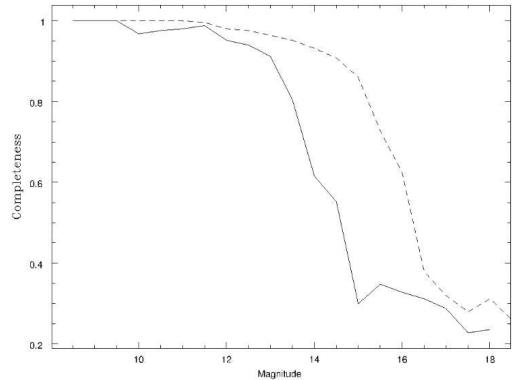}
\figcaption{\label{completeness} Completeness fraction determined by
artificial-star tests for the G305.3+0.2 cluster region ($K$ = solid line, $H$
= dashed line). }
\end{figure}

To provide a robust determination of the KLF and the IMF, we must determine the
completeness of our data, which we established by performing artificial star
tests.  Five artificial stars at a time were inserted into the cluster region;
the small number was chosen to avoid significantly changing the crowding
characteristics.  IRAF-DAOPHOT was then run on the images to determine the
number of artificial stars that were successfully recovered.  The procedure was
repeated 50 times for each magnitude bin ($\Delta m = 0.5$), for a total of 250
artificial stars added in each bin in $H$ and in $K$.
Figure~\ref{completeness} shows the results; completeness falls sharply to
about 25\% at $H \sim 16.5$, $K \sim 15.5$.  We can compare these magnitudes
with the turnover in the ``field luminosity function'', which also probes
incompleteness.  The counts in the field turned over sharply at $K \simeq 16$,
in reasonable agreement with the artificial star estimate of incompleteness.

\subsubsection{The $K$ Luminosity Function}

\begin{figure}
\includegraphics[angle=0,width=9cm]{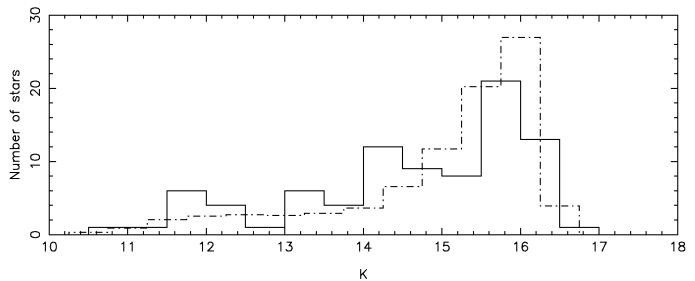}
\figcaption{\label{afgl-KLF-bg} K-band luminosity functions for G305.3+0.2 cluster (solid) and field (dashed),
normalized to the same total number of stars.  Note the peak is shifted to
brighter magnitudes for the cluster.}
\end{figure}

\begin{figure}
\includegraphics[angle=0,width=9cm]{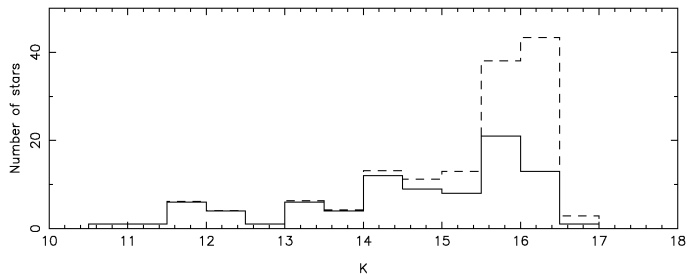}
\figcaption{\label{afgl-KLF-corr}
Completeness-corrected K-band luminosity function for G305.3+0.2
cluster (dashed).  The uncorrected KLF is overplotted (solid) for comparison.}
\end{figure}

Knowing our incompleteness, we can calculate the KLF for the cluster.
Figure~\ref{afgl-KLF-bg} shows the uncorrected data, with the field
``luminosity function'' normalized to the same total number of stars
overplotted for comparison.  Figure~\ref{afgl-KLF-corr} shows the results after
correcting for incompleteness by dividing the number of stars in each magnitude
bin by the recovered fraction of artificial stars.

As expected, there is an
overabundance of bright stars ($K < 14.5$) in the cluster region relative to
the field.  This is not an artifact of incompleteness; the completeness
fraction at this magnitude is $\sim 90\%$, and we expect incompleteness to be
higher in the cluster than the field due to the effects of crowding.  Using the
number counts corrected for field star contamination (as discussed in
\S\ref{g305phot}) and incompleteness, we fit a slope to the number counts in
bins of $\Delta K = 0.5$.  We excluded sources fainter than $K = 15.5$ from the
fit since errors in the incompleteness determination are likely to dominate the
number counts.  We derived a slope of $0.21 \pm 0.06$ for log $N_*$.  This
slope is somewhat flatter than the KLFs derived for more massive embedded
clusters (e.g. $0.41 \pm 0.02$ for NGC 3576 from \citet{figueredo02}, $0.40 \pm
0.03$ for W42 from \citet{blum00}).  This suggests that this cluster is more
weighted toward massive stars than the norm.

\subsubsection{The Initial Mass Function}

In order to better compare our results with the literature, and to explore how
much of a difference the use of multi-color photometry and spectra of the
massive stars make in the determination of the IMF, we used two methods to
derive an IMF for the G305+00.2 cluster.  For both methods we use a distance to
the cluster of 4.0 kpc (as discussed in \S\ref{sec:resafgl}).  The first
IMF-determination method, which uses only the KLF, is commonly employed even
when multi-color photometry and spectra are available
\citep[e.g.][]{figueredo02,blum00}.  This method is simply a transformation
from $K$ magnitude bins to mass bins.  To make this transformation, we first
correct the observed $K$ for distance and extinction as discussed in
\S\ref{sec:resafgl}.  Using the stellar evolutionary models of
\citet{newmodels} for solar metallicity, we relate the mass for each track to
an absolute $K$ magnitude for a star on the ZAMS.  We transformed $L_{bol}$ to
$K$ using the bolometric corrections from \citet{vacca} for the early spectral
types and \citet{malagnini} for later spectral types.  We then use the
intrinsic $V-K$ colors from \citet{bbrett} for A-M stars and from
\citet{wegner} for O and B stars.  Finally we interpolate linearly between the
masses available on the evolutionary tracks to find the masses corresponding to
our magnitude bins, and fit a power law to the resulting mass function.  Our
resulting IMF slope is $\Gamma = -1.5 \pm 0.3$, excluding the two lowest-mass
bins where incompleteness is significant.

Our second method of determining the IMF made use of our multi-color photometry
and spectra to estimate individual extinctions and masses for cluster members.
Spectral typing of the brightest cluster stars allows their mass to be
determined fairly accurately for a given stellar evolutionary model.  For the
models described above, the mass of an O6V star is approximately 40
M$_{\odot}$, that of a B0V star is 15 M$_{\odot}$, and that of a B2V star is 8
M$_{\odot}$.  Although spectra are not available for most of the cluster stars,
their masses, as well as extinctions to the individual stars, can be estimated
from the accurate relative photometry.  The presence of an O supergiant in the
cluster suggests that, while the most massive stars have begun to evolve away
from the main sequence, none have yet gone supernova, and less massive stars
should still be on the zero-age main sequence.  Therefore, with the exception
of the few most massive stars (for which we can estimate masses from their
spectral types) the cluster stars should be scattered around the zero-age main
sequence (ZAMS) primarily by differential extinction and rather than the effects of
stellar evolution.  We can then use the same models and conversions from
theoretical to observed quantities described for the KLF method, with
additional transformations from $T_{eff}$ to $H-K$ using intrinsic colors from
from \citet{bbrett} and \citet{wegner} and from $T_{eff}$ to spectral type from
\citet{repolust} or \citet{johnson66}.

This transformation from theoretical to observed quantities allows us to place
the ZAMS on our CMD.  If the cluster is sufficiently young that we can neglect
the effects of stellar evolution, as discussed in the previous paragraph, we
expect the ZAMS will lie in the middle of the distribution of cluster stars.
The ZAMS derived from the evolutionary tracks of \citet{newmodels} is
overplotted on the distance and extinction-corrected CMD in
Figure~\ref{ZAMSa}. A significant number of stars are bluer than the ZAMS on
this plot.  We interpret these as stars which are less extincted than those
used to determine the average cluster extinction and thus have been
over-corrected by using the mean extinction.  The scatter of stars around the
ZAMS suggests that the extinction varies across the cluster region.  To correct
for this, we move the stars along the direction of the reddening vector until
they lie on the ZAMS.  If the resulting extinction differs from the mean
cluster value by more than $A_V = 5$ for a given star, we exclude the star from
the analysis, as it probably suffers from poor photometry.   Examination of the
color image of the cluster region (Figure~\ref{afglimage}) suggests that the
variation in internal extinction in this region is relatively small; no dust
lanes or color variations across the cluster are visible to the eye.  The exact
value selected for the cutoff is somewhat arbitrary, but does not greatly
affect the results; most of the sources thus excluded have derived extinctions
that differ from the median value by $A_V = 10$ or more.   

Using the positions of the extinction-corrected photometry along the ZAMS, we
are able to more accurately place stars in mass bins.  The endpoints of the
bins were determined by the masses for which theoretical tracks are present in
the models we used.  In order to have an adequate number of stars in each bin
we constructed bins using alternate tracks for the endpoints, rather than every
track.  The analysis was repeated for three different metallicities ($Z=0.1,
0.02, 0.001$) using the evolutionary tracks of \citet[][Z = 0.1]{mowlavi},
\citet[][Z=0.02]{schaller,newmodels} and \citet[][Z=0.001]{schaller}.  For the
solar-metallicity case the high-mass points ($M > 9 M_\odot$) are from
\citet{newmodels} while the lower-mass points are from \citet{schaller}.  The
difference in $K$ for the two solar-metallicity tracks is always less than 0.1
magnitudes for the masses where the two sets of tracks overlap and for most
masses is less than 0.03 magnitudes.  The high metallicity model should be
considered only as a limiting case since such a high metallicity is not
expected.  The use of such a wide range of metallicities allows us to estimate
the importance of this parameter on the final IMF determination.

\begin{figure}
\includegraphics[angle=0,width=9cm]{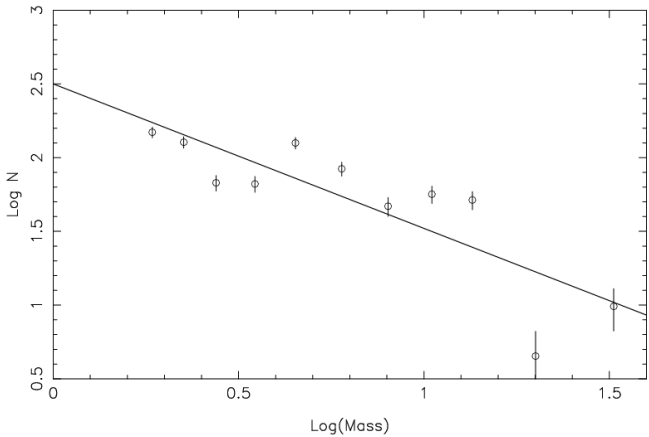}
\figcaption{\label{afgl-IMF} The completeness-corrected IMF for the G305.3+0.2 cluster.  The fitted
  values are the masses derived from the \cite{schaller} stellar evolutionary
  tracks with a distance to the cluster of 3.4 kpc.  The plotted error bars are
  given by assuming the error in the number of stars in a mass bin is equal to
  the square root of the number of stars in the bin.  The fitted line has a
  slope of -0.96.}
\end{figure}

Given these sets of mass bins, for each metallicity we determine the number of
stars per unit logarithmic mass interval after correcting for completeness.  We
then fit a power law to the data.  The two lowest-mass bins ($M < 2
M_{\odot}$), where incompleteness was significant, were excluded from the fit;
uncertainty in the completeness correction applied could significantly
influence the results in these mass bins.  The resulting completeness-corrected
IMF for the cluster is plotted in Figure~\ref{afgl-IMF}.  The solar-metallicity
models yield an IMF slope $\Gamma = -0.98 \pm 0.2$, where the quoted errors are
only the formal fit errors and should be considered an underestimate.  The
low-metallicity tracks yield $\Gamma = -1.01 \pm 0.2$ for the same distance,
suggesting that the cluster IMF determination is insensitive to metallicity for
solar and sub-solar values.  The $Z = 0.1$ tracks give $\Gamma = -0.88 \pm
0.15$.

\subsubsection{Comparison of the IMF Methods}

The IMF slopes we derive using these two methods are marginally consistent
within the error bars: $\Gamma = -1.5 \pm 0.3$ for the KLF method, and $\Gamma
= -0.98 \pm 0.2$ for the CMD + spectroscopy method assuming solar metallicity.
Comparing these results individually to the Salpeter slope would lead to
different conclusions, however.  The KLF method produces a slope that is very
close to the Salpeter value, while the slope from the CMD + spectroscopy method
differs from Salpeter by about 2$\sigma$.  While this difference in slopes
could arise purely from statistical uncertainty, various systematic effects
should cause the KLF-derived slope to be steeper than the CMD-derived slope, as
we observe.  If the more massive stars are preferentially located toward the
center of the cluster, as expected due to mass segregation, and if the
extinction is higher in the center of the cluster, the mean extinction used in
the KLF determination would be systematically low for the more massive stars.
This method would then underestimate the masses the highest mass stars, thus
steepening the slope of the IMF.  Evidence that this effect may be at work is
provided by the six brightest cluster members, all of which lie redward of the
ZAMS in Figure~\ref{ZAMSa} while the fainter members are scattered more evenly.
A difference in $A_K$ (and thus $M_K$) of 0.2 corresponds to 1-2 subtypes for
massive stars and thus to a difference in the derived mass of at least 2
M$_{\odot}$.

However, mass segregation can only provide a partial explanation for the
difference in the IMF slopes; the stars for which we obtained spectra are not
in the very center of the cluster (since crowding in the 2MASS image used to
select spectroscopic targets prevented us from selecting targets in the cluster
core).  An additional possible source of systematic error in the KLF method
relative to the CMD method lies in field star rejection.  In addition to the
statistical field star rejection described in \S\ref{g305phot}, which was done
before any further analysis and thus applies to both methods, the CMD method
has color-based field star rejection.  The CMD method can reject foreground
objects, which due to lower extinction are bluer than cluster objects, as well
as background objects which are redder than the cluster.  The KLF method
includes these objects, which tend to be fainter on average than the cluster
stars (since they are either at a greater distance or are low-mass foreground
stars) and thus finds an artificially high number of low-mass stars.  We find
the use of $K$ photometry alone to derive the IMF is likely to produce an
overly steep IMF in regions with significant field contamination or variable
extinction.

\subsubsection{Comparison With Other Young Stellar Clusters}

Most studies of young star clusters have found an initial mass function
consistent with the Salpeter slope of $\Gamma = -1.35$, generally with
uncertainties of 0.1-0.2
\citep[e.g.][]{figueredo02,massey98,trapezium,okumura00}, including the
extremely massive R136 cluster in the LMC \citep{massey98}.  A review of the
results is provided in \citet{massey03}.  In the case of NGC 6611, reanalysis
of the same data by different authors has produced dramatically different
results; an IMF of $-1.1 \pm 0.1$ was found by \citet{hill93}, while a
reanalysis with different treatment of extinction produced $-0.7 \pm 0.2$
\citep{massey95}, suggesting that the systematic effects are at work in IMF
determinations that are at least as important as the statistical errors, as we
see in this work.  Slopes significantly flatter than Salpeter have been
reported for the Arches cluster near the Galactic Center \citep{figer99},
though later work suggests that this result is an artifact of mass segregation;
\citet{stolte02} found a very flat IMF in the core of the Arches Cluster with a
steeper IMF at larger radii, with an overall slope consistent with a Salpeter
value.  The flatness we observe in both the KLF and the IMF for the G305+00.2
cluster using the CMD + spectroscopy method may similarly be due to mass
segregation.  In addition to the extinction effects mentioned previously,
fainter stars in the outskirts of the cluster could be indistinguishable from
the field star density (especially given the high field star density due to the
location of the cluster in the Galactic plane) and not fall within the cluster
boundaries we employ.

\subsection{The KLF for the G353.4-0.36 cluster}

\begin{figure}
\includegraphics[angle=0,width=9cm]{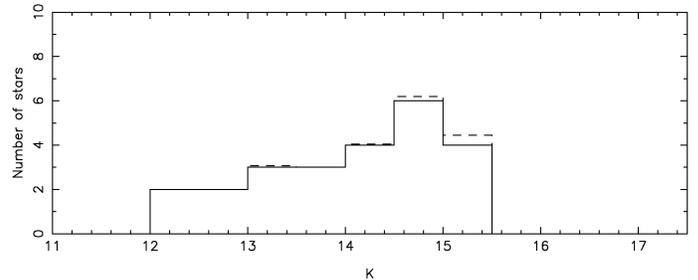}
\figcaption{\label{g353-KLF}
The raw (solid line) and completeness-corrected (dashed line) KLF for the G353.4-0.36 cluster.  
Only sources with $H-K > 1$, which fall in the cluster color sequence, have been included.}
\end{figure}

Completeness tests were performed for the G353.4-0.36 cluster using artificial
stars as discussed above, and the completeness-corrected KLF is plotted in
Figure~\ref{g353-KLF}.  Since the cluster is significantly less crowded and faint
cluster stars less common, our detections in this cluster are nearly complete
in K, even though our detection limit is brighter than in the G305.3+0.2
cluster.  The turnover at $K = 15.5$ appears to be genuine rather than an
artifact of completeness.  Perhaps lower-mass stars in this cluster are still
more deeply embedded in the gas and dust, and thus we observe only the massive
objects.

Due to the small number of stars detected in this cluster ($N = 25$, only 7 of
which were detected in $H$) and to the early evolutionary stage of the objects,
we did not attempt to determine an IMF for this cluster or to place objects on
the ZAMS.  While the individual objects we observed in the G353.4-0.36 cluster
were intriguing and worthy of further study, we cannot analyze the cluster as a
whole because there are so few objects.

This cluster is a very promising target for study at other wavelengths more
suited than the NIR to the study of YSOs and even earlier stages of star
formation; the methanol masers and likely presence of massive YSOs suggest that
several stages of massive star formation can be studied in this region.

\section{Summary}

We present NIR images and spectroscopy of two young stellar clusters near radio
sources G353.4-0.36 and G305+00.2.  Our $K$-band spectrum of the brightest cluster
star in the G305+00.2 cluster show it to be an O5Ib-O6Ib star.  Although the
range of luminosities of supergiants prevents us from determining an exact
distance, this identification suggests a larger distance than radio distance to the nearby
methanol masers \citep{walsh97} of 3.3 kpc.  We also obtained spectra of early
two B stars in the cluster.  There was no nebular emission present in the
G305+00.2 cluster, though a ridge of nebular emission, coinciding with
8~\micron\ emission and masers, is present $\sim$ 1\arcmin\ away and may
indicate sequential star formation, with the masers and gas indicating ongoing
star formation and the cluster the result of earlier star formation.  We
computed the KLF and IMF of this cluster, and found them to be steeper than
that reported for most young clusters ($\Gamma = -0.98 \pm 0.2$ for the more
reliable CMD-based method) but generally consistent with the Salpeter value.
We find that computing the IMF based only on a single color of photometry is
prone to systematic errors when differential extinction and field-star
contamination are significant.

Our $K$-band spectra of two of the three stars we observed in the G353.4-0.36
cluster were featureless, while the other showed CO absorption, which is
consistent either with a cool foreground giant or a YSO.  The absolute
magnitudes derived based on the distance to the radio sources are too bright
for these objects to be solar-mass YSOs.  None of the objects were detected in
our $J$-band photometry, making identification as YSOs based on NIR excess
impossible.  They remain candidate massive YSOs, and observations at other
wavelengths are needed to make a positive identification.   The images of this
cluster showed a region with intense nebular emission embedded in a very dark
cloud where earlier stages of star formation are progressing. 

\clearpage

\begin{deluxetable}{cccccccccc} 
\tablecaption{\label{AFGLews} {\small{Magnitudes and spectral line identifications of
  sources in G305.3+0.2 for which we obtained spectra.  Numbers in parentheses
  in the photometry are the errors in the least significant digit.}}}
\tablewidth{0pt}
\tablecolumns{10}
\tabletypesize{\small}
\tablehead{ 
\colhead{} &
\colhead{RA} & 
\colhead{DEC} & 
\multicolumn{3}{c}{Photometry} & 
\multicolumn{3}{c}{Spectral Properties} &
\colhead{Spectral} \\
\colhead{Star} &
\colhead{13$^{\rm h}~11^{\rm m}$} & 
\colhead{-62$^o$} & 
\colhead{J} & 
\colhead{H} & 
\colhead{K} & 
\colhead{Species} &
\colhead{$\lambda$ ($\mu$m)} & 
\colhead{EW (\AA)} & 
\colhead{Type} 
}
\startdata
A1 &41$^{\rm s}$\hspace{-2pt}.04&32\arcmin\ 56\arcsec\hspace{-3pt}.8 &11.75(1)&10.39(3)\tablenotemark{a}&9.58(3)\tablenotemark{a}&Br$\gamma$& 2.166 & -5.7$\pm$0.6 & O5V-O6V \\ 
&&&&&&\ion{N}{3}& 2.116 & -2.7$\pm$0.7 &	 \\ 
&&&&&&\ion{C}{4}& 2.078 & $\gtrsim$-0.8&  \\  \tableline
A2& 33$^{\rm s}$\hspace{-2pt}.88 &  33\arcmin\ 27\arcsec\hspace{-3pt}.1  
   &12.310(1)&11.02(3)\tablenotemark{a}&10.34(2)\tablenotemark{a} 
  &Br$\gamma$& 2.166 & 6.2$\pm$1.2 & B0V-B1V \\ 
&&&&&& \ion{He}{1}& 2.112 & 0.7$\pm$0.2 &  \\ \tableline
A3& 39$^{\rm s}$\hspace{-2pt}.50 & 33\arcmin\ 28\arcsec\hspace{-3pt}.2  
   &14.063(4) & 12.646(4)  & 11.97(2) 
  &Br$\gamma$ & 2.166 &  5.9$\pm$1.3 & B2V-B3V \\ 
\enddata
\tablenotetext{a}{2MASS magnitude}
\end{deluxetable}

\begin{deluxetable}{llllllll} 
\tablecaption{\label{afgl-phottable} Sample photometry for all stars in G305.3+0.2
  cluster region.  Magnitudes $< 11.5$ are taken from the 2MASS Point Source Catalog,
  since the IRIS2 images are saturated.  Sources A1, A2, and A3 are listed
  first followed by the remaining sources.  Field stars that were removed
  before deriving the luminosity and mass functions are included.}
\tablewidth{0pt}
\tabletypesize{\scriptsize}
\tablecolumns{8}
\tablehead{ 
\colhead{RA (J2000)} & 
\colhead{DEC (J2000)} & 
\colhead{J} & 
\colhead{$\Delta$J} & 
\colhead{H} & 
\colhead{$\Delta$H} & 
\colhead{K} & 
\colhead{$\Delta$K} 
}
\startdata
13:11:41.040 & -62:32:56.77 & 11.751 & 0.010 & 10.394 & 0.027 & 9.575 & 0.029 \\ 
13:11:33.877 & -62:33:27.12 & 12.310 & 0.001 & 11.020 & 0.027 & 10.342 & 0.023 \\ 
13:11:39.503 & -62:33:28.17 & 14.063 & 0.004 & 12.646 & 0.004 & 11.969 & 0.018 \\ 
\tableline
13:11:37.680 & -62:33:09.60 & 11.949 & 0.041 & 10.776 & 0.047 & 10.185 & 0.037 \\ 
13:11:36.286 & -62:33:13.30 & 12.488 & 0.010 & 11.519 & 0.001 & 10.655 & 0.037 \\ 
13:11:41.620 & -62:33:17.40 & 12.936 & 0.029 & 11.650 & 0.038 & 10.953 & 0.034 \\ 
13:11:41.111 & -62:33:18.39 & 16.309 & 0.036 & 14.514 & 0.018 & 11.142 & 0.002 \\ 
13:11:39.268 & -62:33:24.85 & 13.467 & 0.003 & 12.095 & 0.003 & 11.494 & 0.018 \\ 
13:11:39.439 & -62:33:03.63 & 13.218 & 0.003 & 12.018 & 0.002 & 11.524 & 0.018 \\ 
13:11:43.767 & -62:33:26.39 & 16.065 & 0.104 & 13.312 & 0.003 & 11.594 & 0.018 \\ 
13:11:40.045 & -62:33:18.89 & 13.549 & 0.006 & 12.236 & 0.006 & 11.596 & 0.018 \\ 
13:11:38.141 & -62:33:13.66 & 13.428 & 0.003 & 12.270 & 0.003 & 11.745 & 0.018 \\ 
13:11:39.493 & -62:33:10.25 & 16.021 & 0.129 & 14.851 & 0.116 & 11.860 & 0.004 \\ 
13:11:40.458 & -62:33:03.65 & 15.989 & 0.019 & 14.766 & 0.029 & 11.920 & 0.003 \\ 
13:11:40.021 & -62:33:07.26 & 13.884 & 0.004 & 12.591 & 0.003 & 11.970 & 0.018 \\ 
13:11:40.992 & -62:33:07.86 & 14.258 & 0.005 & 12.845 & 0.005 & 12.123 & 0.018 \\ 
13:11:36.748 & -62:33:11.14 & 14.115 & 0.010 & 12.898 & 0.005 & 12.153 & 0.053 \\ 
13:11:34.747 & -62:33:24.02 & 14.067 & 0.010 & 12.742 & 0.002 & 12.311 & 0.029 \\ 
13:11:34.525 & -62:33:11.13 & 14.362 & 0.010 & 12.969 & 0.004 & 12.334 & 0.044 \\ 
13:11:40.031 & -62:33:11.38 & 14.315 & 0.010 & 13.098 & 0.014 & 12.518 & 0.018 \\ 
13:11:40.433 & -62:33:23.29 & 17.628 & 0.108 & 15.615 & 0.049 & 12.698 & 0.005 \\ 
13:11:39.217 & -62:33:08.44 & 14.799 & 0.010 & 13.578 & 0.007 & 12.789 & 0.164 \\ 
13:11:38.080 & -62:32:59.21 & 16.267 & 0.017 & 13.972 & 0.005 & 12.872 & 0.018 \\ 

\enddata
\tablecomments{The complete version of this table is in the electronic
edition of the Journal.  The printed edition contains only a sample.}
\end{deluxetable}

\begin{deluxetable}{lccccc} 
\tablecaption{\label{g353table} Photometric data for the spectroscopic targets in the G353.4-0.36 cluster.}
\tablewidth{0pt}
\tablecolumns{6}
\tablehead{ 
\colhead{ID} & 
\colhead{RA (2000)} & 
\colhead{DEC (2000)} & 
\colhead{J} & 
\colhead{H} & 
\colhead{K} 
}
\startdata
B1 & 	17:30:27.8 &  	-34:41:28.1 & 	\nodata &  14.14 & 	12.85 \\ 
B2 & 	17:30:27.9 & 	-34:41:34.7 & 	\nodata &  14.59 & 	13.47 \\ 
B3 & 	17:30:27.8 & 	-34:41:40 & 	\nodata &  \nodata &  14.38 \\ 
\enddata
\end{deluxetable}

\clearpage

\acknowledgements

We thank the AAT and Chris Tinney for assistance with the IRIS2 instrument.  We
thank Phil Massey, Margaret Hanson, and the anonymous referee for comments that
improved this paper.  This research has made use of the SIMBAD database,
operated at CDS, Strasbourg, France.  This publication makes use of data
products from the Two Micron All Sky Survey (2MASS), which is a joint project
of the University of Massachusetts and IPAC, funded by NASA and NSF.  A.C. was
supported in part by NASA through the American Astronomical Society's Small
Research Grant Program.

\clearpage

\clearpage


\begin{thebibliography}

\bibitem[Argon et~al.(2000)]{masers2} Argon, A. L., Reid, M. J., \& Menten, K. M. 2000, \apjs, 129, 159
\bibitem[Becker et~al.(1994)]{becker} Becker, R. H., White, R. L., Helfand, D. J., \& Zoonematkermani, S. 1994, \apjs, 91, 347
\bibitem[Bessell \& Brett(1988)]{bbrett} Bessell, M. S. \& Brett, J.M. 1988, \pasp, 100, 1134 
\bibitem[Blum et~al.(1997)]{blum97} Blum, R. D., Ramond, T. M., Conti, P. S., Figer, D. F., \& Sellgren, K. 1997, \aj, 113, 1855
\bibitem[Blum et~al.(2000)]{blum00} Blum, R. D., Conti, P. S., \& Damineli, A. 2000, \aj, 119, 1860 
\bibitem[Blum, Damineli, \& Conti(2001)]{blum01} Blum, R. D., Damineli, A., \& Conti, P. S. 2001, \aj, 121, 3149 
\bibitem[Bloom et~al.(2002)]{bloom02} Bloom, J. S. et~al. 2002, \apj, 572, L45
\bibitem[Carey et~al.(2000)]{carey2000} Carey, S. J., Feldman, P. A., Redman, R. O., Egan, M. P., MacLeod, J. M., \& Price, S. D. 2000, \apj, 543, L157
\bibitem[Carpenter(2003)]{carponline} Carpenter, J. 2003, http://www.astro.caltech.edu/~jmc/2mass/v3/transformations/
\bibitem[Caswell et~al.(1995)]{g305masers} Caswell, J. L., Vaile, R. A., \& Forster, J. R. 1995, \mnras 277, 210
\bibitem[Caswell et~al.(2000)]{masers} Caswell, J. L., Yi, J., Booth, R. S., \& Cragg, D. M. 2000, \mnras 313, 599
\bibitem[Chakraborty et~al.(2000)]{chakrabotry} Chakraborty, A., Ojha, D. K., Anandarao, B. G., \& Rengarajan, T. N. 2000, \aap, 364, 688
\bibitem[Chan, Henning, \& Schreyer(1996)]{chan96} Chan, S. J., Henning, T., \& Schreyer, K. 1996, A\&AS, 115, 285
\bibitem[Clark \& Porter(2004)]{clarkporter} Clark, J.S. \& Porter, J.M. 2004,
  \aap, 427, 839
\bibitem[Conti \& Blum(2002)]{blum02} Conti, P. S. \& Blum, R. D. 2002, \apj, 564, 827 
\bibitem[Cotera et~al.(1996)]{coteraarches} Cotera, A. S., Erickson, E. F., Colgan, S. W. J., Simpson, J. P., Allen, D. A., Burton, M. G. 1996, \apj, 461, 750 
\bibitem[Cotera et~al.(1999)]{cotera99} Cotera, A. S., Simpson, J. P., Erickson, E. F., Colgan, S. W. J., Burton, M. G., \& Allen, D. A. 1999, \apj, 510, 747 
\bibitem[Cotera \& Leistra(2005)]{cotera05} Cotera, A. S. \& Leistra, A. L. 2005,  in prep.
\bibitem[Dutra \& Bica(2001)]{db01} Dutra, C. M. \& Bica, E. 2001, \aap, 376, 434 
\bibitem[Dutra \& Bica(2000)]{db00} Dutra, C. M. \& Bica, E. 2000, \aap, 359, 9

\bibitem[Dutra et~al.(2003)]{db03} Dutra, C. M., Ortolani, S., Bica, E., Barbuy, B., Zocalli, M., \& Momany, Y. 2003, \aap, 408, 127
\bibitem[Dutra et~al.(2003b)]{db03a} Dutra, C.M., Bica, E., Soares, J., \&
  Burbuy, B. 2003, \aap, 400, 533 
\bibitem[Elmegreen et~al.(2000)]{elmegreen00} Elmegreen, B. G., Efrernov, Y., Pudritz, R. E., \& Zinecker, H. 2000, in Protostars \& Planets IV: University
  of Arizona Press; eds. Mannings, V., Boss, A. P., \& Russell, S. S., 841
\bibitem[Figer, Morris, \& McLean(1996)]{figer96} Figer, D. F., Morris, M., \& McLean, I. S.  1996, in The Galactic Center: Astronomical Society of the
  Pacific Conference Series; ed Grebel, R., 263 
\bibitem[Figer et~al.(1999)]{figer99} Figer, D. F., Kim, S. S., Morris, M., Serabyn, E., Rich, R. M., \& McLean, I. S. 1999, \apj, 525, 750
\bibitem[Figuer\^{e}do et~al.(2002)]{figueredo02} Figuer\^{e}do, E., Blum, R. D., Damineli, A., \& Conti, P. S. 2002, \aj, 124, 2739 
\bibitem[Forster \& Caswell(2000)]{forster00} Forster, J. R. \& Caswell, J. L. 2000, \apj, 530, 371 
\bibitem[Gardner \& Whiteoak(1978)]{whiteoakcs} Gardner, F. F. \& Whiteoak, J. B. 1978, \mnras, 183, 711
\bibitem[Gardner \& Whiteoak(1978)]{whiteoakh2co} Gardner, F. F. \& Whiteoak, J. B. 1984, \mnras, 210, 23
\bibitem[Greene \& Lada(1996)]{greenelada96} Greene, T. P. \& Lada, C. J. 1996, \aj, 112, 2184 
\bibitem[Hanson \& Conti(1995)]{hanson95} Hanson, M. M. \& Conti, P. S. 1995, \apj, 448, L45
\bibitem[Hanson et~al.(1996)]{hanson96} Hanson, M. M., Conti, P. S., \& Rieke, M. J. 1996, \apjs, 525, 750 
\bibitem[Hanson, Hayworth, \& Conti(1997)]{hanson97} Hanson, M. M., Howarth, I. W., \& Conti, P. S. 1997, \apj, 489, 698
\bibitem[Harju et~al.(1998)]{sio} Harju, J., Lehtinen, K., Booth, R. S., \& Zinchenko, I. 1998, A\&AS, 132, 211
\bibitem[Hillenbrand et~al.(1993)]{hill93} Hillenbrand, L. A., Massey, P., Strom, S. E., \& Merril, M. K. 1993, \aj, 106, 1906
\bibitem[Hillenbrand \& Carpenter(2000)]{trapezium} Hillenbrand, L. A. \&  Carpenter, J. M. 2000, \apj, 540, 236
\bibitem[Huang et~al.(1999)]{huang} Huang et~al. 1999, \apj, 517, 282
\bibitem[Ishii et~al.(2001)]{ishii} Ishii, M., Nagata, T., Sato, S., Yao, Y., Jiang, Z., \& Nakaya, H. 2001, \aj, 121, 3191
\bibitem[Johnson(1966)]{johnson66} Johnson, H. L. 1966, \araa, 4, 193
\bibitem[Lada \& Lada(2003)]{lada03} Lada, C. J. \& Lada, E. A. \araa, 41, 57
\bibitem[Maeder(1981)]{maeder} Maeder, A. 1981, \aap, 101, 385
\bibitem[Malagnini et~al.(1986)]{malagnini} Malagnini, M. L., Morossi, C.,  Rossi, L. \& Kurucz, R. L. 1986, \aap, 162, 140
\bibitem[Massey et~al.(1995)]{massey95} Massey, P., Johnson, K. E., \& DeGioia-Eastwood, K. 1995, \apj, 454, 151
\bibitem[Massey \& Hunter(1998)]{massey98} Massey, P. \& Hunter, D. 1998, \apj, 493, 180
\bibitem[Massey(2002)]{massey02} Massey, P. 2002, \apjs, 141, 81
\bibitem[Massey(2003)]{massey03} Massey, P. 2003, \araa, 41, 15
\bibitem[Meynet \& Maeder(2003)]{newmodels} Meynet, G. \& Maeder, A. 2003, \aap, 404, 975
\bibitem[Morris \& Serabyn(1996)]{morris96} Morris, M. \& Serabyn, E. 1996, \araa, 34, 645 
\bibitem[Mowlavi et~al.(1998)]{mowlavi} Mowlavi, N., Schaerer, D., Meynet, G., Bernasconi, P. A., Charbonnel, C., \& Maeder, A. 1998,  A\&AS, 128, 471 
\bibitem[Nagata et~al.(1990)]{quintuplet} Nagata, T., Woodward, C. E., Shure, M., Phipher, J. L., \& Okuda, H. 1990, \apj, 351, 83
\bibitem[Nagata et~al.(1995)]{nagata95} Nagata, T., Woodward, C. E., Shure, M., \& Kobayashi, N. 1995, \aj, 109, 1676
\bibitem[Oasa, Tamura, \& Sugitani(1999)]{oasa} Oasa, Y., Tamura, M., \&  Sugitani, K. 1999, \apj, 526, 336
\bibitem[Okumura et~al.(2000)]{okumura00} Okumura, S., Atsushui, M., Nishihara, E., Watanabe, E., \& Yamashita, T. 2000, \apj, 543, 799 
\bibitem[Price et~al.(2002)]{price02} Price, P. A. et~al. 2002, \apj, 572, L51
\bibitem[Repolust et~al.(2004)]{repolust} Repolust, T., Puls, J., \& Herrero, A. 2004, \aap, 415, 349
\bibitem[Rieke \& Lebofsky(1985)]{extinction} Rieke, G. H. \& Lebofsky,  M. J. 1985, \apj, 288, 618
\bibitem[Schaller et~al.(1992)]{schaller} Schaller, G., Schaerer, D., Meynet, G., \& Maeder, A. 1992, A\&AS, 96, 269 
\bibitem[Stolte et~al.(2002)]{stolte02} Stolte, A., Grebel, E. K., Brandner, W.,\& Figer, D. F. 2002, \aap, 394, 459 
\bibitem[Vacca, Garmany, \& Shull(1996)]{vacca} Vacca, W. D., Garmany, C. D., \& Shull, J. M. 1996, \apj, 460, 914 
\bibitem[Val'tts et~al.(2000)]{masers3} Val'tts, I. E., Ellingsen, S. P., Slysh, V. I., Kalenskii, S. V., Otrupcek, R. \& Larionov, G. M. 2000, \mnras, 317, 315
\bibitem[Walsh et~al.(1997)]{walsh97} Walsh, A. J., Hyland, A. R., Robinson, G., \& Burton, M.G. 1997, \mnras, 291, 261
\bibitem[Wegner(1994)]{wegner} Wegner, W. 1994, \mnras, 270, 229 
\bibitem[Whiteoak, Otrupcek, \& Rennie(1982)]{whiteoakco} Whiteoak, J. B., Otrupcek, R. E., \& Rennie, C. J. 1982, PASAu, 4, 434
\bibitem[Wilson \& Mezger(1970)]{wilson70} Wilson, T. L, \& Mezger, P. G. 1970, \aap, 6, 364
\bibitem[Wood \& Churchwell(1989)]{wood89} Wood, D. O. \& Churchwell, E. 1989, \apj, 340, 265 
\bibitem[Zinchenko, Henkel, \& Mao(2000)]{zinchenko} Zinchenko, I., Henkel, C., \& Mao, R.Q. 2000, \aap, 361, 1079
\end{thebibliography}
\end{document}